%% file: mainArxiv.tex
\documentclass{article}

\usepackage{arxiv}

\usepackage{natbib}
\usepackage[utf8]{inputenc} 
\usepackage[T1]{fontenc}    
\usepackage{hyperref}       
\usepackage{url}            
\usepackage{booktabs}       
\usepackage{amsfonts}       
\usepackage{nicefrac}       
\usepackage{microtype}      
\usepackage{lipsum}
\usepackage{graphicx}
\graphicspath{{./images/}}
\usepackage{amsmath}

\usepackage{multirow}  
\usepackage{booktabs}   
\usepackage{array}      
\usepackage{adjustbox}  
\usepackage{multirow}   
\usepackage{tabularx}   
\usepackage{xcolor}
\usepackage{soul}
\usepackage{longtable}

\usepackage{tikz}
\newcommand*\circled[1]{\tikz[baseline=(char.base)]{\node[shape=circle,draw,inner sep=1.5pt,fill=black,text=white] (char) {#1};}}

\newcommand*\circledH[1]{\tikz[baseline=(char.base)]{\node[shape=circle,inner sep=0.5pt,fill=gray!40,text=white] (char) {#1};}}

\newcommand*\circledD[1]{\tikz[baseline=(char.base)]{\node[shape=circle,inner sep=0.5pt,fill=blue!40,text=white] (char) {#1};}}

\newcommand*\circledF[1]{\tikz[baseline=(char.base)]{\node[shape=rectangle, minimum size=1.5em, inner sep=0.5pt,fill=blue!40,text=white, draw=none] (char) {#1};}}

\title{GoldMind: A Teacher-Centered Knowledge Management System for Higher Education --- Lessons from Iterative Design}

\author{
 Gloria Fernández-Nieto \\
  Faculty of Information Technology\\
  Monash University\\
  Melbourne, VIC, Australia \\
  \texttt{gloriamilena.fernandeznieto@monash.edu} \\
   \And
  Lele Sha \\
  Faculty of Education\\
  University of Macau\\
  Macau, China \\
  \texttt{lelesha@um.edu.mo} \\
  \And
  Yuheng Li\\
  Faculty of Information Technology\\
  Monash University\\
  Melbourne, VIC, Australia \\
  \texttt{yuheng.li@monash.edu} \\
  \And
  Yi-Shan Tsai\\
  Faculty of Information Technology\\
  Monash University\\
  Melbourne, VIC, Australia \\
  \texttt{yi-shan.tsai@monash.edu} \\
  \And
  Guanliang Chen\\
  Faculty of Information Technology\\
  Monash University\\
  Melbourne, VIC, Australia \\
  \texttt{guanliang.chen@monash.edu} \\
  \And
  Yinwei Wei\\
  School of Software\\
  Shandong University\\
  Jinan, Shandong, China \\
  \texttt{weiyinwei@sdu.edu.cn} \\
  \And
  Weiqing Wang\\
  Faculty of Information Technology\\
  Monash University\\
  Melbourne, VIC, Australia \\
  \texttt{teresa.wang@monash.edu} \\
  \And
  Jinchun Wen\\
  Faculty of Information Technology\\
  Monash University\\
  Melbourne, VIC, Australia \\
  \texttt{jimwendev@gmail.com} \\
  \And
  Shaveen Singh\\
  Faculty of Information Technology\\
  Monash University\\
  Melbourne, VIC, Australia \\
  \texttt{shaveen.singh@monash.edu} \\
  \And
  Ivan Silva\\
  Faculty of Information Technology\\
  Monash University\\
  Melbourne, VIC, Australia \\
  \texttt{ivan.silvaferaud@monash.edu} \\
  \And
  Yuanfang Li\\
  Faculty of Information Technology\\
  Monash University\\
  Melbourne, VIC, Australia \\
  \texttt{yuanfang.li@monash.edu} \\
  \And
  Dragan Gašević\\
  Faculty of Information Technology\\
  Monash University\\
  Melbourne, VIC, Australia \\
  \texttt{dragan.gasevic@monash.edu} \\
  \And
  Zachari Swiecki\\
  Faculty of Information Technology\\
  Monash University\\
  Melbourne, VIC, Australia \\
  \texttt{zach.swiecki@monash.edu} \\
}

\begin{document}

\maketitle
\begin{abstract}
  Designing Knowledge Management Systems (KMSs) for higher education requires addressing complex human–technology interactions, especially where staff turnover and changing roles create ongoing challenges for reusing knowledge. While advances in process mining and Generative AI enable new ways of designing features to support knowledge management, existing KMSs often overlook the realities of educators’ workflows, leading to low adoption and limited impact. This paper presents findings from a two-year human-centred design study with 108 higher education teachers, focused on the iterative co-design and evaluation of GoldMind—a KMS supporting in-the-flow knowledge management during digital teaching tasks. Through three design–evaluation cycles, we examined how teachers interacted with the system and how their feedback informed successive refinements. Insights are synthesised across three themes: (1) Technology Lessons from user interaction data, (2) Design Considerations shaped by co-design and usability testing, and (3) Human Factors, including cognitive load and knowledge behaviours, analysed using Epistemic Network Analysis.
\end{abstract}


\input{sections/1-Introduction}

\input{sections/2-Background}
\input{sections/3-IterativeStudy}

\input{sections/4-Methods}

\input{sections/5-LessonsLearnt}
\input{sections/6-Discussion}

\bibliographystyle{unsrt}  


\bibliography{bibliography}

\end{document}

%% file: sections/1-Introduction.tex
\section{Introduction}

Knowledge Management Systems (KMSs) increasingly use advanced technologies to support educators in digital environments in the context of higher education. For example, process mining can help inform process insights at scale automatically in teachers’ practices \cite{he2021leveraging, ShareFlows2025}. Generative AI can be used to understand user context (e.g., vision models can extract information from images) and make complex information more accessible, for instance, by summarising lengthy documents \cite{JARRAHIHuman-AIPartnership}. These capabilities are particularly relevant in education, where knowledge in educational practices is highly contextual, distributed across platforms, and challenging to capture and transfer without disrupting daily tasks \cite{ShaIUI2025, Fernandez-Nieto2024}.

In higher education, the ability to effectively capture and reuse institutional knowledge is vital for universities. This is particularly the case in contexts with high staff turnover, the prevalence of sessional academics, and evolving teaching responsibilities \cite{LiuPD2020, LeeTurnover2010}. A well-designed KMS promises to support the continuity of teaching practices, foster organisational learning, and enable the transfer of expertise from experienced educators to newcomers. However, implementing such systems in educational settings poses several challenges, particularly when technology, design, and human factors are not adequately integrated.

Technologically, many existing KMSs in education have focused on isolated knowledge management processes — such as storing or accessing content — without addressing knowledge management holistically across its four core processes: access, organisation, sharing, and application \cite{ishak2010integrating, alavi2001knowledgeProcesses}. This fragmented design limits the systems' capacity to support educators in dynamic, real-world teaching environments.
For example, a recent literature review by Fan
and Beh \cite{FAN2024KSharing} found limited empirical research on knowledge-sharing tools in higher education, with minimal input from educational stakeholders, suggesting a disconnection between the tools provided and the actual technological needs of educators 
Similarly, Bilquise and Shaalan \cite{Bilquise2023} found that although knowledge sharing and creation are the most researched knowledge management processes in universities, 
the technological support for collaborative practices and organisational culture, such as teachers knowledge workflows, remains underdeveloped. For example, Yun et al. \cite{yun2025generative}, in a non-university context, highlights the need to implement mechanisms that enable knowledge flow and support the connection between existing knowledge and users’ needs.
Yet, such tools fell short in addressing complex knowledge flows and routine teaching needs. These shortcomings illustrate persistent \textbf{technological challenges}, including the lack of intelligent, context-aware features (e.g., adaptive interfaces, personalised recommendations, or workflow integration) that support effective knowledge management for educators.

To overcome limitations in existing KMSs, a growing body of research advocates for \textbf{human-centred approaches} to KMS \textbf{design} — i.e., approaches that position educators not merely as end users but as co-designers of systems that reflect their needs, behaviours, and values \cite{HCDJoseph2014, Sarka2019KMTI}. Human-Centred Design (HCD) emphasises meaningful participation and iterative refinement based on user input. In the context of education, where teachers work under cognitively demanding and time-constrained conditions, systems that are not carefully designed with the user in mind risk adding cognitive burden or disrupting established practices \cite{GalgotiaKMHigherEducation2022}.
These challenges underscore the importance of integrating user feedback throughout the design process to ensure long-term adoption and usability.

Alongside technological and design considerations, \textbf{human factors} — such as motivation, trust, and cultural attitudes toward knowledge sharing — play a critical role in the success of KMSs \cite{rehman2021empirical}. As emphasised by Sarka et al. \cite{Sarka2019KMTI}, future research must more thoroughly address the dynamic interplay between human factors and technology. Their work identifies several key dimensions that shape engagement with knowledge management systems in educational settings: behavioural --- e.g., how technology influences teaching and knowledge management practices, cultural---- e.g., openness to sharing knowledge, and experience related --- e.g., differences in digital literacy across expert groups. These factors are especially relevant in higher education, where teachers operate within complex, cognitively demanding, and time-constrained environments \cite{rehman2021empirical, Bilquise2023}. However, KMSs are often developed without a clear understanding of teachers’ real-world workflows \cite{rehman2021empirical}, resulting in systems that do not align with daily teaching practices or the cognitive demands educators face \cite{corso2009CoPProcesses, GalgotiaAI-KMS2021}. Introducing systems that increase cognitive load or fail to align with existing workflows can discourage usage and reduce long-term impact \cite{GalgotiaAI-KMS2021}. To be effective, KMSs must adapt to users’ real needs, routines, and working approaches — not the other way around.

Despite these known concerns, the technological, design, and human factors involved in implementing KMSs in higher education settings — particularly through iterative, human-centred approaches and within the context of real-world teaching tasks — remain underexplored. Exploring these factors can offer insights into how KMSs can be effectively integrated into educators’ workflows, how users engage with such technological support, and how these systems influence knowledge management behaviours over time. In response to this gap, our study presents the design and evaluation of 
\emph{GoldMind}, a KMS developed through an iterative, human-centred process in collaboration with 108 university teachers over two years. The system supports the capture, storage, sharing, and reuse of expert knowledge in authentic teaching contexts. Through a series of design and evaluation cycles, we examined how teachers interacted with the system and how it supported their knowledge management practices. 

Here, we  synthesise lessons learnt across three system iterations, highlighting key insights across three interrelated themes:
(1) \textit{Technology}: We evaluated user interaction metrics (e.g., task time, failure rates) to assess system \textit{effectiveness} and guide improvements.
(2) \textit{Design}: Through co-design sessions and usability testing, we iteratively refined system features and interface elements based on educator feedback.
(3) \textit{Human Factors}: We analysed users’ cognitive load, perceptions, and qualitative feedback to evaluate how the system aligned with educators’ work practices.
To complement this analysis, we also explored knowledge management behaviours using epistemic network analysis (ENA) to model user interaction logs, uncovering patterns in knowledge management sub-processes such as storing, sharing, and applying knowledge.


By articulating these findings, this paper contributes: (i) practical guidance for designing teacher-centred KMSs based on one of the largest iterative design and evaluation studies with authentic teachers, and (ii) introduces GoldMind, a KMS that supports in-the-flow knowledge capture and dissemination in higher education.




%% file: sections/2-Background.tex
\section{Related Work}

In this section, we first provide an overview of knowledge management in education. To contextualise the discussion, we draw on insights from  recent systematic literature reviews and related works. 
These includes reviews on knowledge management in higher education \cite{DIVAIO2021220, Asiedu31122022, Bilquise2023, BemMachado04032022}, future directions in knowledge management research \cite{manesh2020knowledge}, and emerging knowledge management topics in information technologies \cite{farooq2024review, yun2025generative}. We synthesise findings from these works 
to identify key technological, design, and human factors 
challenges that must be addressed to ensure that knowledge management practices in higher education, supported by KMSs, remain relevant and impactful.

\subsection{Knowledge Management for Education}

Knowledge management plays a pivotal role in capturing and disseminating knowledge within organisations, especially in knowledge-intensive fields like education \cite{LiuPD2020}. In universities, where staff turnover is often high — due to retirement, institutional transfers, staff teaching reassignments, or the appointment of sessional academics and teaching assistants — preserving institutional knowledge is essential \cite{LeeTurnover2010,figueron2015influences,sharma2022conceptual}. Doing so likely mitigates knowledge loss and ensures continuity in teaching and learning practices \cite{arun2005360,ravikumar2022impact}.

Effective knowledge management enables the \textbf{access}, \textbf{capture}, \textbf{storage}, \textbf{sharing}, \textbf{dissemination}, and \textbf{application} of knowledge—processes essential for informed decision-making, problem-solving, and innovation \cite{Carroll2003, ishak2010integrating, Sangeeta2015, Suazo2024, BemMachado04032022}. Capture, storage, and access form the foundation for knowledge management activities; sharing allows individuals and teams to replicate effective practices; dissemination and application ensures that knowledge is widely spread and used to enhance performance and organisational growth \cite{du2007role, al2018development, al2018impact}.

Bose and Sugumaran \cite{Bose2003} and Wong and Aspinwall \cite{Wong2004} outline knowledge management processes that encompass the flow of knowledge within an organisation — from capture to dissemination and application — providing a useful framework for defining knowledge management-related activities. Experts engage in these processes both tacitly and explicitly in their daily workflows \cite{DAGENAIS2020101778,Asiedu31122022}. In educational contexts, knowledge management processes are particularly critical for addressing staff turnover, supporting organisational learning, and enabling effective decision-making  Asiedu
et al. \cite{Asiedu31122022}, Rehman et al. \cite{rehman2021empirical}.

Educators engage in knowledge management activities \cite{Bose2003, rehman2021empirical, Bilquise2023} such as navigating repositories (\textit{access}), uploading and organising materials (\textit{storage}), annotating and collaborating on documents (\textit{sharing}), and applying institutional knowledge (e.g., policies or best practices) to complete teaching tasks (\textit{application}). Understanding these processes offers a structured method to capture and disseminate expertise, especially from experts to novices \cite{
rehman2021empirical, ShareFlows2025}.

Drawing on established knowledge management frameworks \cite{
andreeva2011knowledge, Carroll2003, ishak2010integrating, Rubenstein-Montano2015, LiuPD2020}, and motivated by the documented gap in effectively capturing, storing---and especially---disseminating knowledge to mitigate loss and avoid unnecessary disruptions \cite{Debs2023}, our study focuses on the design and implementation of a KMS that supports both knowledge management capture and dissemination processes. To address the challenges posed by staff turnover in universities, we designed features in a KMS that allow teachers to capture knowledge (e.g., document relevant information and workflows) while performing their regular tasks. The system also enables the dissemination of both newly captured and existing knowledge (e.g., pushing recommendations), in ways that enable its reuse by others, thus promoting its application and helping complete the knowledge flow within higher education institutions. 

\subsection{Technological challenges}

Given the potential of knowledge management processes to enhance educational practice, researchers have explored how technology can support educators’ knowledge management activities. Kaplan and Haenlein \cite{Kaplan2014CoPEducation} demonstrated how collaborative tools such as wikis — with messaging, content integration, and announcement features — can facilitate knowledge sharing. Similarly, email and other communication tools have long been used to support collaboration and information exchange among educators \cite{hinton2003knowledge}. Carroll et al. \cite{Carroll2003} developed a tool combining video conferencing, chat, email, and notebooks to help rural educators capture and organise their resources, activity plans, and student outputs for future reuse. More recently, Arpaci et al. \cite{MOOCSTAM2020} evaluated MOOCs as integrative KMSs and found them useful for storing and disseminating knowledge, though with varying degrees of effectiveness depending on the user context.

Generative AI (GenAI) is also creating new opportunities for knowledge management, although most existing research in this area has not focused specifically on educational contexts. For example, Brynjolfsson et al. \cite{brynjolfsson2025generative} demonstrated the use of a GenAI-based conversational assistant in the workplace to enhance the productivity of customer support agents, particularly less experienced workers, by improving task performance, and enhancing customer interactions. Similarly, in an organisational context, Alavi et al. \cite{alavi2024knowledge} illustrated how GenAI can influence knowledge management processes. Their study highlights GenAI’s potential in knowledge storage and retrieval, enabling rapid access to large knowledge bases and reshaping how employees access to knowledge and interact with KMS.

Despite these advances, the implementation of knowledge management technologies in higher education remains limited due to persistent challenges that complicate their integration into educational settings. Higher education institutions are inherently knowledge-intensive and multidisciplinary, which increases the complexity of knowledge and makes it difficult for existing technologies to support efficient capture, retrieval and reuse of knowledge 
\cite{GalgotiaAI-KMS2021, Asiedu31122022}. Organisational and technological decisions are often made without adequately considering educators’ real needs \cite{corso2009CoPProcesses, Fernandez-Nieto2024}. Although GenAI models can be tailored with domain-specific datasets to adapt to specific domains, they often fail to incorporate human voices in their training processes \citep{weisz2024design}. This may cause the lack of trust in AI-generated recommendations \citep{RODGERS2023122373, alavi2024knowledge}. In addition, Galgotia and Lakshmi \cite{GalgotiaAI-KMS2021} emphasised the absence of mature, intelligent (e.g., AI-powered) KMSs in higher education settings that align with real educators' needs. Furthermore, Galgotia and
Lakshmi \cite{GalgotiaAI-KMS2021} emphasised the lack of mature, intelligent (e.g., AI-powered) KMSs in higher education that truly meet the needs of educators. Galgotia and Lakshmi \cite{GalgotiaKMHigherEducation2022} argued that KMSs are rarely embedded in educators’ everyday practices, resulting in fragmented knowledge flows and avoidable barriers. Building upon the work by Brynjolfsson
et al. \cite{brynjolfsson2025generative}, Yun et al. \cite{yun2025generative}, which explores the impact of generative technologies on productivity in non-educational domains (e.g., project management), such integration has the potential to enhance educator productivity, particularly for novice teachers. KMSs are essential to manage large, dynamic knowledge bases and provide timely, context-sensitive support (e.g., recommendations \cite{brynjolfsson2025generative}). Specifically for GenAI, there is a documented need to balance human-derived tacit knowledge with AI-generated explicit knowledge \cite{alavi2024knowledge}. Additionally, the risks of using GenAI to disseminate sensitive information remain underexplored \cite{alavi2024knowledge}. In this paper, we directly address these challenges and discuss the lessons learned during the design and implementation of a KMS for a higher education institution.

\subsection{Design Challenges}

Key design challenges of KMSs include fragmented support for knowledge processes \cite{MOOCSTAM2020}, limited integration of human-centred design, methodological evaluation gaps \cite{manesh2020knowledge, pascalHCDKM_2013, Sarka2019KMTI, horbanproblem2022}, and a lack of context-specific design tailored to educational settings \cite{Uchidiuno2023}, all of which hinder KMSs adoption \cite{manesh2020knowledge}.

One promising direction to address these challenges is the integration of a holist knowledge management process approach within established technology adoption frameworks. For example, Arpaci et al. \cite{MOOCSTAM2020} extended the Technology Acceptance Model (TAM) to incorporate knowledge management processes — such as access, storage, sharing, and application — to better evaluate the role of MOOCs in supporting knowledge management. While this work demonstrated that knowledge storage and access were positively perceived, the findings also underscore the need for deeper exploration of user experiences and perceptions to understand the impact of implementation.

Human-Centred Design (HCD) offers an important pathway toward developing more effective KMSs. HCD involves the iterative participation of users, designers, and researchers in co-developing tools that align with real-world needs \cite{Louis1992, HCDJoseph2014, Krol2025CHI}. As stated by Manesh et al. \cite{manesh2020knowledge}, human-centered approaches to designing KMSs approaches help mitigate resistance to change and help to unleash true potential of people, by understanding their goals, and particular needs. 
For example, Nguyen Ngoc et al. \cite{HCD_KnowledgeManagement2022} reviewed the opportunities of using HCD in KMSs, particularly as a means of incorporating human voices to support variety and customisation in system design.
Similarly, Touré et al. \cite{Tour2014ReDesigningKM} conducted focus groups to redesign a poorly adopted knowledge management tool. 
In a non educational context, HCD has also been explored in the context of knowledge management for manufacturing activities \cite{Fakhar2021}.
In another example, Pascal et al. \cite{pascalHCDKM_2013} created a knowledge management portal for partner identification in R\&D, using a science-based design approach (i.e., inspired by research evidence) combined with human-centered design. However, their approach relied more on literature and application reviews than on participatory design. In the work by Almunawar and Ordñez de Pablos \cite[Chapter~4]{DigitalisationChapter4-2022}, online surveys were used to investigate university teachers’ perceptions of knowledge management technology use in Asia. While the study captured teachers’ views, it provided only descriptive statistics and lacked qualitative insights into users’ knowledge management preferences, behaviours, and needs.

Our study extends this body of work by involving both individual teachers and teaching teams — i.e., staff employed by higher education institutions in active teaching and coordination of university courses — in the co-design and evaluation of the GoldMind KMS. The system was developed to integrate seamlessly into their workflows and address the challenges identified by Sarka et al. \cite{Sarka2019KMTI}, including gaining a holistic understanding of educators' needs and the mechanisms required to support them. We conducted co-design sessions—a core human-centred design (HCD) approach that fosters equal participation between users and design experts in the development of digital tools \cite{wojciechowski2023evaluating}. This method supports the creation of culturally and contextually relevant systems by incorporating the lived experiences and perspectives of intended users \cite{Uchidiuno2023}.

Guided by prior research on knowledge management processes and technologies \cite{Sarka2019KMTI, horbanproblem2022, GalgotiaKMHigherEducation2022}, our study gathered qualitative evidence from educators about their current knowledge management practices, tools, and needs. In addition, our iterative design approach allowed us to observe how teachers interacted with GoldMind, helping us identify further opportunities for improvement. This process ensured that human factors — such as working approaches, values, and social dynamics — were embedded in the system’s design.

\subsection{Human Factors Challenges}

According to Sarka et al. \cite{Sarka2019KMTI}, research on knowledge management technologies needs to account for the complex interplay between human actors and technological systems. Their work highlights behavioural (e.g., how technology changes practice), cultural (e.g., openness to sharing), and generational (e.g., differing digital literacy or expertise) factors that influence knowledge management behaviours in education. 

A deep understanding of user communities’ needs and workflows is critical for designing usable KMSs and for developing policies that foster their adoption among educators. Successful knowledge management also requires recognising the authentic authors of knowledge — namely, communities of practice \cite{Carroll2003}. This becomes especially important when incorporating emerging technologies such as GenAI, where intellectual property and ethical considerations are paramount \cite{alavi2024knowledge}. In line with this, understanding how teaching practices should inform the design of KMSs -- to support both individual and collective needs, such as reducing cognitive load -- and how these systems, in turn, shape teachers’ practices, remains an open and underexplored area in the knowledge management literature \cite{Sarka2019KMTI, horbanproblem2022}.


As highlighted in the literature review by Manesh et al. \cite{manesh2020knowledge}, future avenues for KMS research should include a deeper exploration of quantitative approaches -- such as the use of process modelling languages -- to better understand how knowledge management functions and to support its application within organisations. Beyond tool design, however, challenges also remain in understanding and supporting individual \textbf{knowledge management behaviours}. Gaining a deeper understanding of how educators share, store, access, and apply information can offer valuable insights into whether they lack awareness or structured approaches to their knowledge management behaviours, particularly as revealed through their interactions with KMS tools. This gap underscores the need for KMSs and processes that more closely align with users’ real-world behaviours, practices, and motivations.

\subsection{Contribution to HCI and Research Question}


The current study contributes to the body of research in human–computer interaction (HCI) by addressing the underexplored intersection between KMSs and HCD in higher education. While KMSs have long been positioned as tools for improving organisational learning and knowledge transfer 
\cite{su141168782022, Muthmainah_2023}, their design has often been shaped by technological capabilities rather than educators’ lived experiences. Recent innovations — such as the integration of Process Mining and GenAI \cite{Nannini2025,Somboonpatarakit2024} — present new opportunities to capture and reuse institutional knowledge more effectively \cite{manesh2020knowledge, GalgotiaAI-KMS2021}. Yet, current implementations in educational settings frequently focus on isolated knowledge management processes such as storage or access \cite{Kaplan2014CoPEducation, 
horbanproblem2022}, often overlooking the complexity of real-world teaching workflows and organisational knowledge flows \cite{GalgotiaKMHigherEducation2022, corso2009CoPProcesses}.

Human-centred approaches offer a promising pathway for addressing these limitations. Prior studies have shown that engaging users in iterative design can increase system relevance, usability, and adoption \cite{HCDJoseph2014, Tour2014ReDesigningKM, horbanproblem2022, HCD_KnowledgeManagement2022}. However, most research has remained limited to technology evaluations or design proposals rather than examining how educators actually engage with KMS over time. By involving 108 teachers in the co-design and iterative evaluation of the GoldMind system, our work contributes empirical insights into how educators’ knowledge management behaviours — such as access, storage, sharing, and application — unfold in authentic teaching contexts \cite{DAGENAIS2020101778, alavi2001knowledgeProcesses, GalgotiaKMHigherEducation2022, GalgotiaAI-KMS2021}.

Motivated by the persistent technological, design, and human factors challenges outlined in the literature \cite{Sarka2019KMTI, MOOCSTAM2020, Carroll2003}, we pose the following research question (RQ):

 \begin{itemize}
     \item[\textbf{RQ}] \emph{What technological, design, and human factors challenges emerge through an iterative, human-centred implementation of a KMS in a university setting?}
 \end{itemize}

This question reflects the need to advance HCI research on KMSs by surfacing design lessons and behavioural insights that can guide the development of more context-aware, participatory, and sustainable KMSs in education.
 

%% file: sections/3-IterativeStudy.tex
\section{Iterative Study}

\begin{figure}[h]
\centering
\includegraphics[width=1\textwidth]{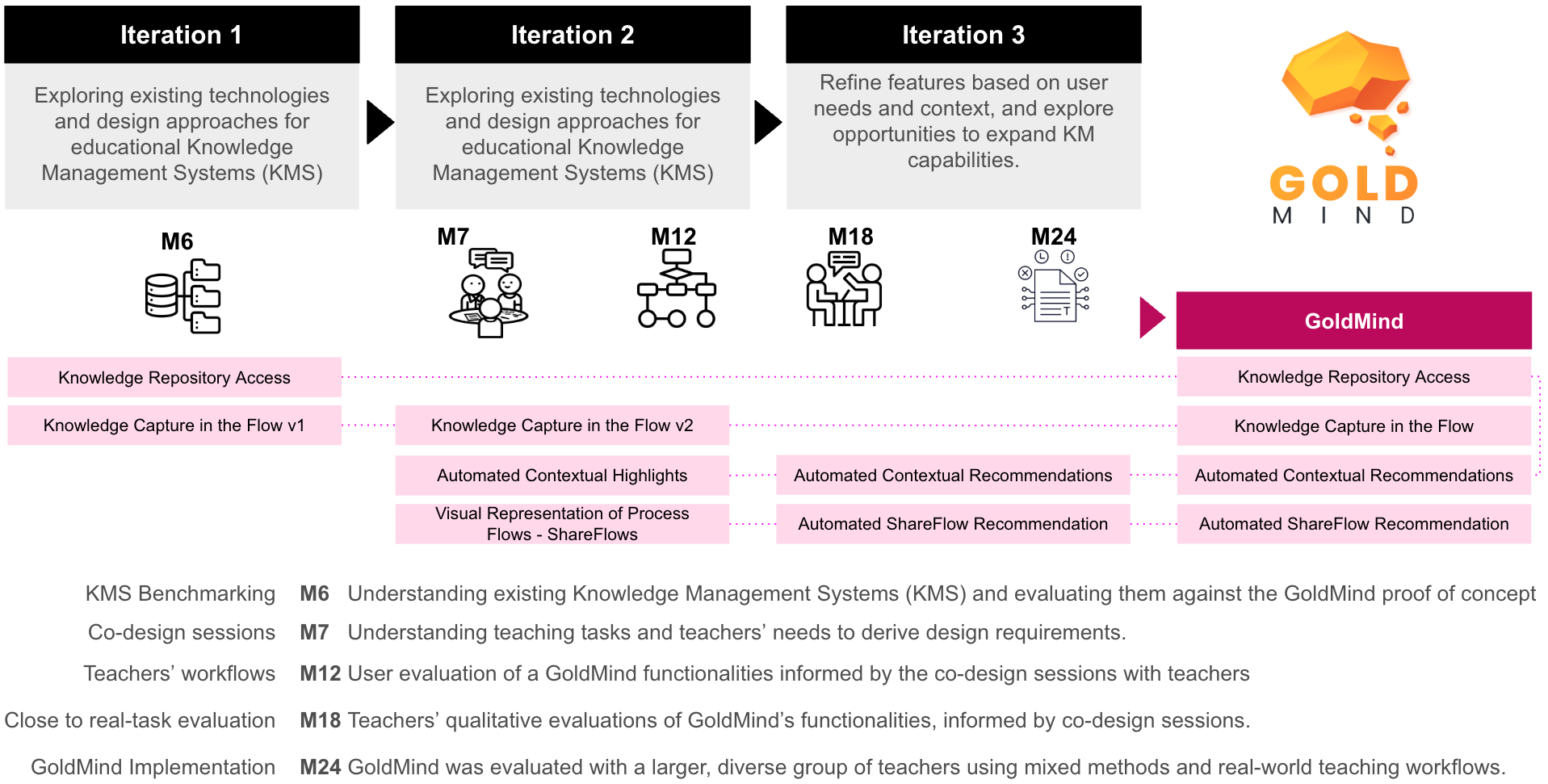}
\caption{
    Human-centred iterative design and implementation of the GoldMind Knowledge Management System. Each iteration involved targeted activities, beginning with benchmarking existing KMSs (M6), followed by co-design sessions to understand teachers' workflows and tasks (M7), user evaluations (M12), and qualitative assessments in near-authentic contexts (M18). The final implementation and large-scale evaluation occurred at M24. Across these phases, features evolved incrementally, including Knowledge Repository, Knowledge Capture in the Flow (v1 to v2), Contextual Highlights, ShareFlow visual representations, and automated recommendation features.
} 
\label{fig:iterations}
\end{figure}

\subsection{Context}

We present a study that follows a human-centred approach \cite{BuckinghamShum2019}. Figure \ref{fig:iterations} shows the three iterations reported in this study, which involved interactions with educators from a research-focused university and an interdisciplinary research team (comprising experts in HCI, learning analytics, storytelling, education, and AI). These collaborations led to the co-creation of GoldMind, a KMS designed to support teachers in seamlessly capturing and disseminating knowledge. GoldMind is accessible via a browser extension\footnote{https://colam.kmass.cloud.edu.au/
}. While the co-design and development of specific GoldMind features are beyond the scope of this paper, further details including technical aspects can be found in other sources \cite{ShaIUI2025, Fernandez-Nieto2024}. 
This paper focuses on the iterative evaluations and involvement conducted with teachers.

Given the complexity of the teaching context and the need for a KMS to support it, 
we adopted an iterative, HCD approach involving authentic teaching teams and real teaching tasks. Our aim was to gather teachers’ feedback and experiences in ecologically valid settings that reflect their actual teaching practice. Each iteration was intended to incrementally improve the design based on evaluation results and teacher feedback \cite{DasIncrementalAgile}. We engaged educators in various instructional roles — including course coordinators, course lecturers, and tutors — to ensure a diverse range of perspectives across the iterations.

We used realistic task scenarios to maintain ecological validity 
in an experimental control setting
\cite{ChiragTaskBasedEvaluation}. This design allowed participants to engage with authentic teaching situations without the risk of affecting live systems (e.g., unintended changes to student marks in the Learning Management System). These tasks were embedded in teaching contexts where educators' expertise was essential for delivering high-quality instruction. The specific context teaching tasks is introduced in the following section.


\subsubsection{Teaching tasks}
\label{sec:teachingTasks}

\begin{figure}[ht]
\centering
\includegraphics[width=1\textwidth]{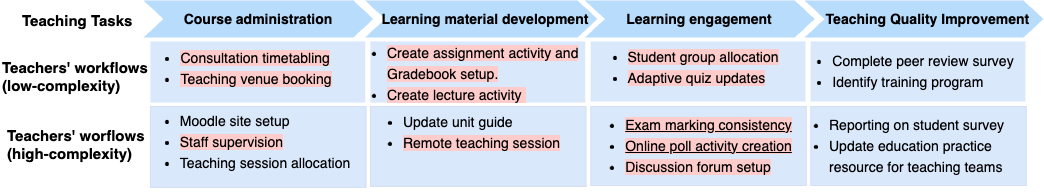}
\caption{Representative teaching tasks performed by higher education teachers in their teaching workflow. 
\textbf{Red} indicates those tasks that are performed more than once per semester. We underscore the tasks selected for evaluation in this study. Task complexity is measured by the time and the number of resources required to complete the task.} 
\label{fig:teachingtasks}
\end{figure}

Teachers, individually or in groups, regularly perform complex tasks to support their courses, such as \textit{creating lecture activities}, \textit{ensuring exam marking consistency}, or \textit{updating teaching resources for teams}. These tasks often involve \textbf{workflows}, including navigating multiple information systems and accessing various resources. For instance, marking requires high levels of consistency, particularly in team settings, to ensure fairness and uphold student outcomes. Without proper guidance, novice teachers tasked with these responsibilities risk introducing inconsistencies. Similarly, managing teaching materials or integrating activities to foster student engagement further add to the complexity of teaching tasks \cite{Fernandez-Nieto2024}.

The already demanding nature of teaching is further complicated when handovers result in the loss of expert knowledge. Over time, senior teachers develop deep expertise in their teaching, such as identifying effective strategies to engage students. However, teacher turnover — due to factors such as promotions or new job opportunities — often leaves insufficient time for documenting the regular workflows of experts \cite{ ShaIUI2025}. This creates difficulties for novice teachers, who must navigate procedures and locate resources independently, a challenge exacerbated in universities where information is scattered across different systems.


Based on the co-design sessions described by Fernandez-Nieto et al. \cite{Fernandez-Nieto2024} 
and evaluations of teachers' workflows reported by Sha et al. \cite{ShareFlows2025},  
we identified 19 representative teaching workflows across four categories: Course Administration, Learning Material Development, Learning Engagement, and Teaching Quality Improvement (see Figure \ref{fig:teachingtasks}). In these co-design sessions, teaching teams — including Chief Examiners, Lecturers, and Tutors\footnote{Chief Examiners are responsible for course coordination, Lecturers manage course delivery and coordination, and Tutors support course management, marking, and tutorial delivery} — collaboratively mapped their workflows using web-based collaborative tools (e.g., Miro). Through affinity diagramming, we systematically analysed and categorised these workflows, providing insights into the nuanced practices of expert teachers and the complexity of their tasks. This understanding informed the design of \textit{Visual Process Representations}, which aim to simplify and communicate these workflows effectively. The design process and its outcomes are detailed in the following sections.

\subsection{Study iterations and participants}
\label{sec:iterationsParticipantsTasks}

\begin{table}[ht]
    \centering
    \caption{Summary of study iterations, participant roles, knowledge management tasks, and task types. Tasks were selected with a focus on knowledge capture and dissemination, aligning with the nature of our design. Task types are abbreviated as KC (Knowledge Capture) and KD (Knowledge Dissemination)}
    \label{tab:iteration_participants}
    \renewcommand{\arraystretch}{1.2} 
    \begin{adjustbox}{max width=\textwidth}
    \begin{tabular}{|p{1.2cm}|c|p{1cm}|p{3.5cm}|p{2.3cm}|p{7.2cm}|}
        \hline
        \textbf{Iteration} & \textbf{Month} & \textbf{Number} & \textbf{Roles} & \textbf{Number of tasks} & \textbf{[Type] Teaching Tasks} \\
        \hline
        \multirow{2}{*}{1} 
        & M6
        & N=15 & Tutor = 9, CE=3, Lecturer=3
        & 3 KC, 3 KD 
        & [Learning engagement] Create a feedforward poll (KC). \newline
        [Learning engagement] Adaptive Moodle quiz update (KC). \newline
        [Learning material development] Update and publish Unit Preview (KC).
        \newline
        [Learning material development] Add learning objectives and materials to a Moodle page. (KD).
        \newline
        [Learning material development] Create a new assignment in Moodle with a Turnitin draft submission (KD).
        \newline
        [Learning material development] Enrol students into Gitlab groups (KD). 
        \\
        \hline
        \multirow{4}{*}{2} 
        & M7
        & N=13
        & Co-design \newline Tutor = 4, CE=4, Lecturer=5 
        & NA 
        & NA \\
        \cline{2-6}
        & M12
        & N=40 \newline
        & Tutor = 26, CE=5, Lecturer=9 & 6 KC, 6 KD & All M6 tasks \newline
        [Teaching quality improvement] Marking correction documentation  for consistency (KC).  \newline
        [Teaching quality improvement] Reporting and documenting a workshop issue via a video (KC).  \newline
        [Course Administration] Creating Miro boards and making them accessible for future use (KC). \newline
        [Learning material development] Set up a scheduled Zoom lecture session (KD). \newline
        [Learning material development] Set up a Moodle forum for a specific unit (KD). \newline
        [Teaching quality improvement] Validate teacher induction (KD). \\
        \hline
        \multirow{4}{*}{3} 
        & M18
        & N=10 & Tutor = 7, CE=1, Lecturer=2 & 3 KC/KD & 
        [Learning material development] Editing gradebook item. \newline
        [Learning engagement] Create a feedforward poll. \newline
        [Learning material development] Set up a scheduled Zoom lecture session. \\
        \cline{2-6}
        & M24
        & N=30 & Tutor = 19, CE=8, Lecturer=3 & 3 KC/KD  & 
        [Learning material development] Editing gradebook item. \newline
        [Learning engagement] Create a feedforward poll. \newline
        [Learning material development] Set up a scheduled Zoom lecture session. \\
        \hline
    \end{tabular}
    \end{adjustbox}
    
\end{table}


The design and implementation of GoldMind followed three iterations. Figure \ref{fig:iterations} provides specific details on each iteration. The first iteration (Month 6 -- \textbf{M6}), conducted in the second semester (late July to December) of 2022, focused on (i) understanding the knowledge management systems available in a higher education institution and (ii) benchmarking existing KMSs against an initial prototype. The second iteration (Month 12 -= \textbf{M12}), conducted in 2023, involved (i) a participatory design process with authentic teaching teams to understand teachers' needs and identify design requirements for educational KMS (March-April 2023) and (ii) prototyping within the research team and evaluating GoldMind features with educators as they completed teaching tasks (August 2023). Finally, the third iteration, conducted in the last semester of 2023 - \textbf{M18} (late July to December) and the first semester \textbf{M24} (late February to mid June) of 2024, i) refined GoldMind’s features and ii) gathered additional qualitative feedback from educators using the KMS to support their teaching tasks. Participants involved in the co-design sessions took part in iterations one and two.

Table \ref{tab:iteration_participants} presents the number of participants per iteration (108 in total), participants' teaching roles, and the tasks that were tested in each of the iterations with the features implemented. All participants were linked to a university but were part of different Departments, including Medicine, Nursing and Health Sciences, and Information Technology. In each iteration, we evaluated GoldMind using the teaching tasks described in Section \ref{sec:teachingTasks}. In line with knowledge management foundations \cite{al2018impact, ishak2010integrating}, effective knowledge management allows the capture and dissemination of knowledge to be used when needed \cite{Rubenstein-Montano2015}. Thus, the iterations considered \textit{Knowledge Capture} (KC) tasks and \textit{Knowledge Dissemination} (KD) tasks. 

For each iteration, this study examines three main groups of factors, outlined below:

\begin{itemize}
    \item The \textbf{technological factors} refer to the challenges, constraints, and opportunities in developing and implementing KMS for teachers. These includes aspects such as knowledge capture, dissemination (retrieval), system automation, and usability enhancements.
    \item The \textbf{design factors} encompass usability, interface design, and interaction mechanisms required to improve teachers engagement with the KMS. These emphasise iterative design refinements, participatory approaches, and context-aware interfaces.
    \item The \textbf{human factors} consider the individual, collaborative, and behavioral dimensions of adopting KMS in higher education institutions. These includes how teachers engage with the system (e.g., perception, adoption, time on task and quality, and interaction), their cognitive load (e.g., mental effort involved in using the KMS), collaboration and communication (e.g., how the system supports knowledge sharing among educators), and knowledge management behaviors (access, storage, sharing, and application). 
\end{itemize}

\subsection{GoldMind Features}
\label{sec:features}
This section outlines the main features developed in each iteration to support knowledge management in teaching tasks. These features were informed by knowledge management frameworks and focus on two core processes: \textit{knowledge capture} and \textit{knowledge dissemination}. The goal is to ensure that knowledge is effectively captured, stored, and shared, while also guiding how institutions implement knowledge management practices.

\subsubsection{Features -- Iteration 1} \label{sec:SystemIteration1} This iteration aimed to evaluate the existing KMSs at a higher education institution and compare them to the first prototype of GoldMind, which included the following features. 

\begin{figure}[t]
    \centering
    \includegraphics[width=1\linewidth]{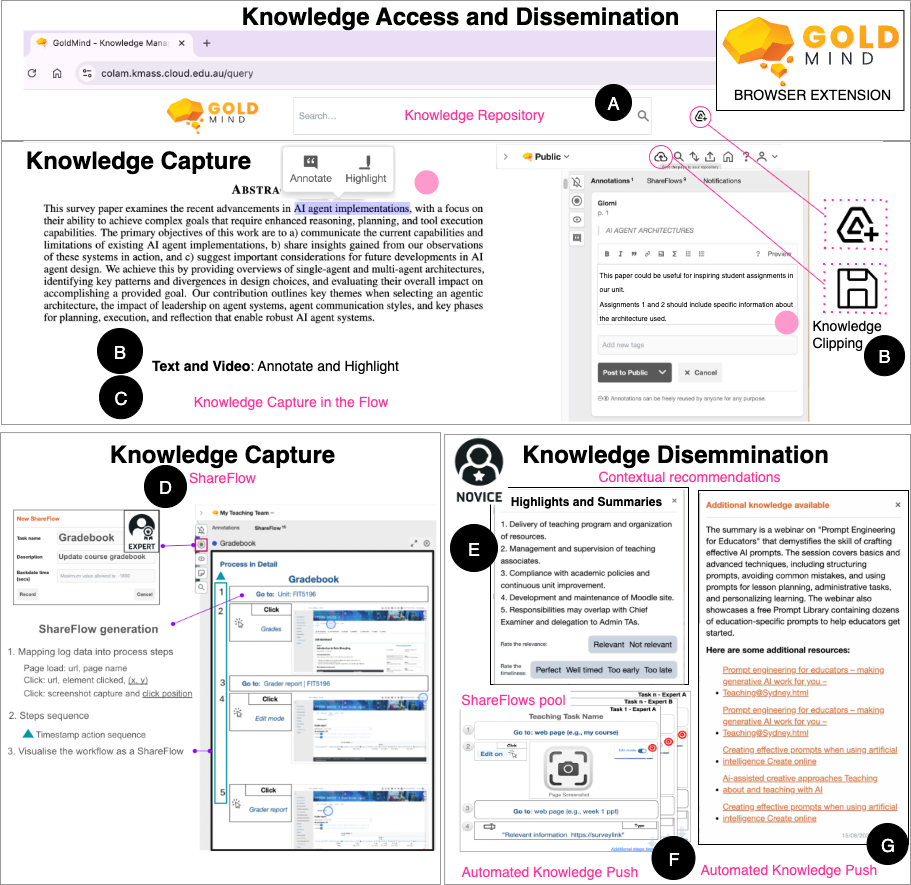}
    \caption{GoldMind Features. The diagram illustrates key components that support knowledge capture and dissemination within the GoldMind platform. Features include: (A) a centralised Knowledge Repository; (B) in-the-flow Knowledge Capture, including knowledge clipping, annotations, and uploading personal knowledge via a browser extension; (C) video annotation capture; (D) Visual Representation of Process Flows through ShareFlows; (E) contextual recommendations powered by highlights and summaries; (F) automated ShareFlow push recommendations; and (G) automated contextual recommendation mechanisms that surface relevant knowledge at the point of need.}
    \label{fig:FeaturesGoldMind}
\end{figure}

\textbf{Knowledge Repository}: \label{sec:featureKBackground} 
This feature supports \textit{knowledge dissemination} by enabling the storage and organisation of knowledge. It indexes existing resources — such as documents and videos from various repositories — for easy retrieval. It uses a Differential Search Index (DSI) model to represent content as dense vector embeddings, enabling efficient document retrieval based on query similarity \cite{Dreano2023EmbedLlamaUL, DSIchen2023TextRetrieval}. This aims to help educators find relevant content within a platform that integrates information from multiple information systems (e.g., Learning Management System and internal university resources for educators, such as TeachHQ) that were previously siloed  (Fig. \ref{fig:FeaturesGoldMind}--\circled{A}).

\textbf{Knowledge Capture in the Flow}: This feature was designed to help teachers \textit{capture knowledge} at varying levels of detail. It aims to enable them to annotate, highlight, and store knowledge while working on regular teaching tasks. To support this, we extended the capabilities of the open web annotation tool \textit{hypothes.is}\footnote{https://web.hypothes.is}. In addition to capturing content from the web, our implementation allows teachers to upload personal and team knowledge from external repositories such as \textit{Google Drive}, integrating it into the knowledge background. We refer to this functionality as \textit{knowledge clipping} (Fig. \ref{fig:FeaturesGoldMind}--\circled{B}).

\subsubsection{Features -- Iteration 2}
\label{sec:SystemIteration2}
Features of Iteration 2 were informed by \textit{co-design sessions} conducted with authentic teaching teams who teach and coordinate subjects in real-world university settings \textit{\cite{Fernandez-Nieto2024}}. Teachers identified the need for capturing granular data, which aligned with features from Iteration 1, and the need for accessing knowledge while working on teaching tasks. 

\textbf{Knowledge Capture in the Flow}: To complement the \textit{knowledge capture} capabilities, we implemented \textit{Video Annotation} by extending \textit{hypothes.is}, which was originally designed for annotating web pages. Given that videos are commonly used by teachers when planning and preparing their classes, the ability to annotate such resources was considered useful for them. This feature allows teachers to capture knowledge in the form of annotations at multiple levels of granularity and in different modes (e.g., text and video annotations) (Fig. \ref{fig:FeaturesGoldMind}--\circled{C}).

\textbf{Visual Representation of Process Flows - ShareFlows}: \label{sec:featureShareFlow} To support expert teachers in \textit{capturing knowledge} — identified in the co-design sessions as a key pain point — we designed, implemented, and evaluated a mechanism to record their task-related trace data. This feature enables experienced educators (\textbf{experts}) to document task steps seamlessly, which can later be proactively recommended to \textbf{novice} teachers via push notifications. In this iteration, the capture process was embedded within teachers’ regular workflows to minimise disruption. The resulting visualisations was designed to support \textit{knowledge dissemination} and were named \textit{ShareFlows} \cite{ShaIUI2025}. The design of ShareFlows applies \textit{data storytelling} principles \cite{Schulz2013, Dykes2015, Ryan2016, Knaflic2017, Echeverria2018, Alhadad_2018}  to present complex expert trace data in a clear, accessible format for novices. The ShareFlow design included: (a) a descriptive title, (b) reduced visual clutter, (c) visual cues (e.g., click highlights) to emphasise key actions, and (d) a scrollytelling layout, a visual storytelling technique, that allows users to navigate steps by scroll up or down a page, enhancing engagement and readability \cite{Scrollytelling2023} (Fig. \ref{fig:FeaturesGoldMind}--\circled{D}). The functionality to recommend ShareFlows to novice teachers during task completion was implemented in Iteration 3.

\textbf{Contextual Highlights}: To support \textit{knowledge dissemination} we implemented contextual recommendations to address teachers' need for access to more relevant and situational knowledge. Using a generative language model (LLaMA v3.1), we developed a feature that summarises key information from documents accessed by teachers, listing relevant insights drawn from existing knowledge sources. When teachers explored a knowledge repository, the system provided recommendations in the form of \textit{highlights}, identifying crucial information within the document (e.g., web page) as pop-up windows (Fig. \ref{fig:FeaturesGoldMind}--\circled{E}). 

\subsubsection{Features -- Iteration 3}
\label{sec:SystemIteration3}
For Iteration 3, we focused on refining mechanisms for disseminating knowledge proactively. The refinements aimed to facilitate knowledge to be access and use automatically, reducing the effort required from teachers to search for information. 

\textbf{Automated ShareFlow Recommendations}. This feature ensures knowledge capture and dissemination occur seamlessly within teachers' workflows, minimising manual effort and disruptions. GoldMind captures novice teachers' trace data about actions they take (e.g., page views, mouse moves, clicks) in real-time to provide timely recommendations by proactively pushing relevant ShareFlows from expert teachers (see Section \ref{sec:featureShareFlow}). To extract key steps from process trace data generated through teachers’ interactions with GoldMind, this feature employs the heuristic miner algorithm \cite{weijters2011flexible}, a widely used process mining algorithm known for its robustness against noisy traces, which are common in human process data \cite{de2012multi}. In essence, GoldMind records detailed task traces from experts during their daily workflows, encodes them as visual step-by-step processes, and proactively recommends them to novices performing similar tasks (Fig. \ref{fig:FeaturesGoldMind}--\circled{F}).

\textbf{Automated Contextual Recommendation}. Leveraging the Llama LLM (version 3.1), we identify the content of the web pages where teachers are working and recommend relevant resources from the knowledge repository. These recommendations are tailored to the task and content being reviewed by the teacher. The LLM generates a synthetic query based on the page content, which is then used to retrieve related resources from the existing Knowledge Repository (see Section \ref{sec:featureKBackground}). These recommendations are deliver as pop-up windows containing a list of resources useful for a particular task (Fig. \ref{fig:FeaturesGoldMind}--\circled{G}).

%% file: sections/4-Methods.tex
\section{Methods}

\subsection{Iterations with authentic teaching teams and tasks}
\label{sec:iterativestudies}

Designing and implementing a KMS requires insights derived from realistic studies and task-based experiments with human participants to understand what works well and what needs refinement \cite{IngwersenIREvaluation}. For complex systems — such as those involving contextual recommendations in information retrieval — it is critical to evaluate how effectively the system supports task completion and goal attainment \cite{ChiragTaskBasedEvaluation}. Both holistic and component-level evaluations are essential for building functional KMS and generating actionable insights for iterative improvement \cite{MOUSTAFA2021107755}.

Using authentic users in iterative studies enhances ecological validity, as it better reflects real-world use compared to evaluations with simulated end-users \cite{rooksby2013wild}. This approach also uncovers practical challenges that educators face when engaging with KMS tools in everyday teaching contexts.

To further balance realism with experimental control, we used real teaching tasks and adopted a simulated task-based approach \cite{ChiragTaskBasedEvaluation}. Simulated tasks allowed participants to engage with the GoldMind system in realistic teaching scenarios without the risks associated with live environments (e.g., accidental changes to student marks in the Learning Management System). The teaching tasks used in this study were elicited through co-design sessions with educators \cite{Fernandez-Nieto2024} 
and are detailed in Section \ref{sec:teachingTasks}, with specific tasks per iteration introduced in Section \ref{sec:iterationsParticipantsTasks}.

Given GoldMind’s complexity, its evaluation involved benchmark test collections and overall metric measurements within teaching tasks. This section outlines the metrics used in each iteration to inform continuous system improvement. Moreover, combining iterative studies with authentic, realistic teaching tasks provides valuable insights that inform successive system redesigns (e.g., \citep{spence2019seeing,taylor2013leaving,vsabanovic2014designing}). Across three iterations, we examined the key challenges of designing and deploying a KMS in higher education contexts. These challenges were analysed through three interrelated lenses: (1) technological factors, (2) design considerations, and (3) human factors.

\subsubsection{Task design} Inspired by the notion of evaluating authentic tasks, we selected evaluation tasks for the GoldMind tool based on the following criteria: (1) tasks that are independently performed using the university’s teaching resources (e.g., intranet) and learning management system (e.g., Moodle), (2) tasks that are complex and time-consuming, and (3) tasks that are frequently performed and therefore highly relevant to teachers’ daily workflows.

Based on these criteria, we selected the evaluation tasks summarised in Table \ref{sec:iterationsParticipantsTasks}, aiming to reflect diverse teaching contexts. Two researchers (co-authors) conducted prior observational work to document educators’ typical workflows and resources. We also invited three expert teachers to perform the tasks while recording their interactions. Insights from these sessions informed the design of realistic and meaningful evaluation tasks.

\subsection{Sources of evidence}
\label{sec:evidences}

Table \ref{tab:sources_of_evidence} summarises the sources of evidence collected from teachers through system evaluations, interviews, and surveys. These sources informed the lessons learned and guided the results presented in the following sections. In each iteration, we explored the three key factors, technological, design, and human. However, it was only in the third iteration that we examined knowledge management behaviours using teachers’ trace data.

\begin{table}[h]
    \centering
    \renewcommand{\arraystretch}{1.2}
    \begin{tabular}{c|c}
        \toprule
        \textbf{Iteration} & \textbf{Source of evidence} \\
        \midrule
        1 (M6) & Tool evaluation metrics (between-subject design) and SUS and NASA surveys 
        (N=15) 
        \\
        \midrule
        2 (M7 and M12) & Interview (co-design) (N=13) \\
        \cline{2-2}
        & Tool evaluation metrics (within-subject design) and SUS and NASA survey (N=40) 
        \\
        \midrule
        3 (M18 and M24) & Tool evaluation metrics (within-subject design) and individual interviews (N=10) \\
        \cline{2-2}
        & Tool evaluation metrics (within-subject design), SUS survey, and individual interviews (N=30) \\
        \bottomrule
    \end{tabular}
    \caption{Summary of interactions, sources of evidence, and explored themes and behaviours.}
    \label{tab:sources_of_evidence}
\end{table}

In the first iteration of the study (M6), teachers were asked to complete six teaching tasks (see Table \ref{sec:iterationsParticipantsTasks}). They were divided into two conditions: Baseline and GoldMind. The baseline condition involved the existing information systems used at the higher education institution. After completing the tasks, participants in each condition were asked to evaluate the usability of the tool(s) using the System Usability Scale (SUS) \cite{lewis2018system}, and to assess cognitive load by using the NASA Task Load Index (NASA-TLX) \cite{HART1988139}. This iteration aimed to identify initial technical challenges in existing systems, evaluate a prototype of GoldMind with early features, and gather insights into human factors to inform future design improvements.

In the second iteration (M12), we conducted a co-design session with authentic teaching teams to identify critical pain points and validate design requirements for the system. An updated prototype of GoldMind was evaluated using the same usability and cognitive load measures from the first iteration.

In the final iteration, a refined version of GoldMind was designed and implemented for evaluation. For comparative purposes, metrics were also collected for a Baseline condition, which was based on the Claude LLM. The Baseline system used a Retrieval-Augmented Generation (RAG) technique to ingest domain-specific data (e.g., university policies or course material) and support teachers in completing teaching tasks via a chatbot interface, which was akin to that of ChatGPT or other contemporary LLM-based chatbots. This Baseline was referred to as ChatUI for evaluation purposes.

The goal of this iteration was to assess the effectiveness of GoldMind in supporting teaching tasks compared to an LLM-based approach. To evaluate usability, we asked the participants in both conditions to complete the adapted version of the SUS survey that was developed to evaluate KMSs. The adaption expanded the original SUS 10 questions into 25 questions covering four key knowledge management practices of knowledge access, storage, sharing, and application, as reported in the knowledge management literature \cite{Carroll2003, ishak2010integrating, Sangeeta2015, Fernandez-Nieto2024}. Additionally, semi-structured interviews were conducted at two time points (M18 and M24) to explore teachers’ perceptions based on the following questions:
i) \textit{How helpful was the information provided by the Baseline/GoldMind for completing the tasks?}
ii) \textit{What features of the Baseline/GoldMind tool did you find most useful?}
iii) \textit{Did you encounter any challenges using the Baseline/GoldMind tool? If so, please describe them.}

\subsubsection{Procedure} 

The tool evaluation was conducted in an experimentation room where participants worked independently on teaching tasks while being observed by a researcher from the adjacent room through a one-way mirror. Each participant used a laptop with an additional screen: the laptop displayed task instructions, while the external screen was used to perform and record tasks via the Chrome browser. Participants were first briefed on the study and given five minutes to explore the tool. The assigned tasks per iteration are detailed in Table \ref{tab:iteration_participants}.

\subsection{Analysis}
\label{sec:analysis}

Given the multiple iterations and sources of evidence used to evaluate GoldMind, the following sections detail the specific analyses and data collected for each iteration, and how they align with the technological \circledF{T}, design \circledF{D}, and human factors \circledF{H}. 

\subsubsection{Iterations 1 and 2} 
\textbf{Tool evaluation metrics}. We adopted a user interaction approach based on trace data to understand how users engage with the system, identify areas for improvement, and inform design decisions \cite{HussainUXTraceDataPerceptions, MeeiTraceData}. In Iteration 1, metrics were primarily derived from manual video annotations, while in Iteration 2, data capture was automated. The Baseline for comparison was the university’s existing knowledge management systems (e.g., intranet repositories, Google Drive). Metrics focused on two task types: (i) dissemination—measuring the time spent searching for task-related information to assess the system’s effectiveness in supporting information dissemination, and (ii) knowledge capture—measuring time spent capturing information (e.g., note-taking for unit planning). The evaluation metrics included:
\begin{itemize}
    \item (i) Retrieval speed: Time taken by the system (Baseline or GoldMind) to return a response to a query \circledF{T}.
    \item (i) Relevant responses ratio: Teachers rated the usefulness of responses on a 0–4 scale. This metric also included the number of actions (e.g., clicks and scrolls) performed to find relevant information \circledF{H}.
    \item (i) Time to action: Time taken by users to initiate the next action after receiving a response (e.g., creating an assignment in Moodle) \circledF{H}.
    \item (i, ii) Task duration: Total time spent on the task \circledF{H}.
    \item (ii) Knowledge capture efficiency: The average cost of capturing knowledge, calculated using time on task (for capture tasks), time spent capturing knowledge per condition, and task completion quality \circledF{T}.
    \item (i) Just enough responses: Number of queries needed to retrieve useful information for completing the task \circledF{T}. 
    \item  Task completion ratio: proportion of tasks completed by participants per condition \circledF{H}.
    \item (i, ii) Task completion quality: Task performance rated on a scale from 0 to 1, based on a predefined rubric \circledF{H}.   
\end{itemize}

For time to action, task duration, knowledge capture efficiency, and just enough responses, we tested for statistical differences between conditions using the mixed-effects model in Formula \ref{equa:completion_timeIt1}, with a fixed effect for condition (\textit{c}) and a random effect for participant (\textit{p}).

\begin{equation}\label{equa:completion_timeIt1}
    Y_{\text{time, duration, capture, responses}} = \beta_0 + \beta_1 \times \text{c} + \gamma_{\text{p}} + \epsilon
\end{equation}

We used a mixed-effects regression model to assess statistical differences in retrieval speed, task completion time, and quality metrics (dissemination and capture) across the two experimental conditions. The model included fixed effects for condition (\textit{c}) and task (\textit{t}), and a random effect for participant (\textit{p}); see Formula \ref{equa:completion_quality}.

\begin{equation}\label{equa:completion_quality}
    Y_{\text{quality}} = \beta_0 + \beta_1 \text{c} + \beta_1 \text{t}+ \gamma_{\text{p}} + \epsilon
\end{equation}

For task completion and the relevant responses ratio, we used a binomial logistic regression model to assess statistical differences between the two conditions (GoldMind vs. Baseline). We analysed the relationship between the condition (independent variable, denoted as c) and the dependent variables (completion and relevance); see Formula \ref{equa:completionrelevance}.

\begin{equation}\label{equa:completionrelevance}
\log\left(\frac{p}{1 - p}\right)= \beta_0 + \beta_1 \text{c}
\end{equation}

We adopted a usability questionnaire to gain quantitative insights into teachers' perceptions of the tools. We used the \textbf{SUS} survey\cite{brooke1996sus} \circledF{D}, which elicited relevance scores based on participants’ perceptions of how useful the system was in supporting the task context (e.g., \cite{Vlachogianni27052022}). Responses were collected on a 5-point Likert scale (1 = strongly disagree, 5 = strongly agree). The numerical values of these responses were used to calculate a final usability score according to the standard SUS scoring procedure \cite{lewis2018system}.

\textbf{NASA-TLX} was used to measure cognitive load across six dimensions \circledF{H}: Mental Demand, Physical Demand, Temporal Demand, Performance, Effort, and Frustration \cite{rubio2004evaluation}. Responses were collected using a 5-point Likert scale (1 = Very Low, 5 = Very High). The numerical values were used to calculate the mean score for each dimension and were visualised using box plots. 

\subsubsection{Iteration 3}
\label{sec:analysisIteration3}
For this iteration, we followed a within-subjects design and simplified the evaluation metrics. The main focus was to assess the system’s effectiveness in making pre-existing knowledge (e.g., captured by experts) accessible to new teachers via automated push dissemination (see features in Section \ref{sec:SystemIteration3}). A chatbot tool based on the Claude LLM served as the Baseline. We also gathered qualitative insights through post-task interviews with teachers.

\textbf{Tool evaluation metrics.} The set of metrics considered for this iteration are as follows:
\begin{itemize}
    \item Reduction in time to access knowledge: We measured the time-on-task for each participant and evaluated how long they spent locating relevant information to complete the assigned tasks \circledF{T}.
    \item Failure rate: During the evaluation, teachers were given a limited amount of time to complete the tasks. We then recorded how many tasks each teacher successfully completed within the allotted time \circledF{H}.
    \item Task completion quality: We tracked user actions—such as clicks, scrolling, and navigation steps—performed while searching for the correct response. These actions were used to assess the efficiency and focus of participants' task completion \circledF{H}.
\end{itemize}

For task completion time and reduction in knowledge management time, we measured statistical differences between conditions using the mixed-effects regression model defined in Formula \ref{equa:completion_timeIt1}. 
For the failure rate, we used the binomial logistic regression model described in Iteration 1.

For participants' task quality, we used fixed effects for condition, start condition (i.e., using Baseline tool first or GoldMind tool first, denoted as \textit{s}) and task (denoted as \textit{t}), considering the interactions between condition (i.e., baseline vs. GoldMind) and start condition, as well as interactions between condition and task, and a random effect for participants, see Formula \ref{equa:completion_qualityIt3}.

\begin{equation}\label{equa:completion_qualityIt3}
\begin{split}
    Y_{\text{quality}} = & \, \beta_0 + \beta_1  \text{c} + \beta_2  \text{s} + \beta_3  (\text{c} \times \text{s}) + \beta_4 \text{t}
    + \beta_5  (\text{t} \times \text{c}) + \gamma_{\text{p}} + \epsilon
\end{split}
\end{equation}

\textbf{Knowledge management SUS survey} \circledF{D}. We adopted a usability questionnaire to gain quantitative insights into participants' perceptions of the tool.  The questionnaire was adapted from the most widely used standardized questionnaire (i.e., SUS) for the assessment of perceived usability \cite{lewis2018system}. The adaption expanded the original 10 questions into 25 questions covering four key knowledge management practices of knowledge access, storage, sharing, and application, as reported in the knowledge management literature \cite{Carroll2003, ishak2010integrating, Sangeeta2015, ShaIUI2025}. 

\textbf{Teachers' Interviews}. To analyse the three four-ended questions described in Section \ref{sec:evidences}, one of the authors, with a background in HCI, coded the transcripts of the responses and used audio recording data to support interpretation. The coding process was guided by predetermined categories outlined in the study procedure (Section \ref{tab:sources_of_evidence}): (i) educators' perceptions of the usefulness of GoldMind and the Baseline \circledF{D} and (ii) educators preferred features of GoldMind and the Baseline \circledF{T}, and (iii) educators' perceptions of the challenges they faced using GoldMind and the Baseline \circledF{H}. Since the categories were distinct, none of the quotes overlapped, and the coder categorised each teacher's quotes into each of the categories above (i-iii). The resulting coded statements were then reviewed by the authors, who engaged in several discussions to select instances that best illustrated the usefulness, preferred features, and challenges of GoldMind and Baseline.

As part of the human factor exploration \circledF{H}, we conducted \textbf{Epistemic Network Analysis (ENA)} \cite{Shaffer_Collier_Ruis_2016} to analyse teachers’ knowledge management behaviours while using the systems (Baseline and GoldMind). In brief, ENA creates network models for each unit of analysis. The nodes of these networks are codes applied to the data---i.e., labels that describe important themes or actions that the units take; the edges represent the co-occurrence of these codes within pre-define windows of data for that unit. Thicker, more saturated edges indicate higher frequency co-occurrences between codes. In this way, ENA can model the patterns of interconnected actions that individuals take. These networks can be analysed using two coordinated representations: (i) a network graph (as described above), and (ii) a summary statistic derived from a dimensional reduction on set of networks in the data. Networks and their summary statistics (called ENA scores) are visualised in the same low-dimensional space where the dimensions are linear combinations of the original co-occurrence variable that explain the maximal amount of variance between the units of analysis. The dimensions can be interpreted using the network diagrams, which are arranged such that edges on the extremes of the dimension correspond the co-occurrences that distinguish units of analysis the most. Both ENA scores and ENA networks can be aggregated to compare subpopulations within the data and networks for different subpopulations can be subtracted to show the co-occurrences that were stronger in one subpopulation compared to the other. Standard statistical techniques can be applied to the ENA scores to test for systematic differences between subpopulations. For more details on ENA, see \cite{Bowman2021}. This analysis was only performed for Iteration 3, as GoldMind and the Baseline were mature enough at that stage to collect teachers’ trace data.

The ENA model was configured as follows: \textit{units} corresponded to participants by task and condition; \textit{codes} represent knowledge management sub-processes (see Table \ref{tab:ENAcodes}); \textit{conversations} include the system and task name (Task 1 or Task 2); \textit{window} size was set to $40$ lines, manually derived by sampling $50$ actions in the trace data and using the median window length; and \textit{metadata} included task outcome (Pass or Fail) and condition order. We constructed the ENA model using the means rotation (MR) approach, which identifies the network embedding space whose first dimension maximises between-group variance. Here, we maximised the difference between the Baseline and GoldMind participants. To visualise the differences between conditions, we generated a subtracted network between the mean network of the GoldMind participants and the mean network of the Baseline participants. This graph showed the connections that occurred more frequently in one condition compared to the other. We complemented this network comparison with statistical tests using a mixed-effects regression model with position on the means-rotated dimension (\textit{MR1}) as the outcome variable, condition (\textit{c}), task (\textit{t}), task order (\textit{s}), and task outcome (\textit{o}) as fixed effects, and participant (\textit{p}) as a random effect, see Formula \ref{equa:ENAMixed-effects}. 

\begin{equation}\label{equa:ENAMixed-effects}
    Y_{\text{MR1}} = \beta_0 + \beta_1 \text{c} + \beta_1 \text{t} + \beta_1 \text{s} + \beta_1 \text{o} + \gamma_{\text{p}} + \epsilon
\end{equation}

To code the data, we identified sequences in the teachers’ trace data and automatically\footnote{using process variation analysis and implemented with the pm4py Python library} 
mapped them to knowledge management sub-processes — i.e., knowledge access, storage, sharing, and application — in alignment with established knowledge management frameworks \cite{Carroll2003, ishak2010integrating, Sangeeta2015, Fernandez-Nieto2024}. The teachers’ trace data used to identify the codes included actions (e.g., clicks, type, navigation, or uploading of new pages) and other metadata, such as URLs, to validate the resources accessed by teachers. Table \ref{tab:ENAcodes} summarises the code names, knowledge management sub-processes and definitions, illustrations of mapped sequences from trace data, and examples of what these may represent in a task context. 

\begin{longtable}[t]
{p{2.7cm} p{4.2cm} p{3.5cm} p{3.5cm}}
\toprule
\small 
\textbf{Code name} & \textbf{KM Process - Definition} & \textbf{Sequences} & \textbf{Example} \\
\midrule
AP.1 Search and Explore 
& Access process: The user searches for a page, explores it, and moves to a new resource. & ['Navigation', 'Scroll', 'Click', 'Navigation'] & Searching for information across multiple web pages. \\
\addlinespace
AP.2 Switch Tabs & Access process: The user navigates across tabs in the browser. & ['Navigation', 'Navigation', 'Navigation'] & Searching for the correct tab where information is located. \\
\addlinespace
AP.3 Read/Engage with Content & Access process: The user interacts with a single resource for a period of time. & ['Click', 'Click'] & Engaging with a page, likely reading relevant information. \\
\addlinespace
AP.4 Open and Interact with Content & Access process: The user opens a new resource and starts interacting with it. & ['Navigation', 'Click'] & Looking for information on a newly opened page. \\
\addlinespace
SP.1 Type and Interact 
& Store process: The user interacts with a resource and types information for a period of time. & ['Click', 'Click', 'Click', 'Click', 'Type'] & Creating a forum in Moodle. \\
\addlinespace
SP.2 Select and Type
& Store process: The user selects an option from a form on a page. & ['Select', 'Type'] & Selecting an option from a form. \\
\addlinespace
SP.3 Type-driven Content 
& Store process: The user types in a single resource for a period of time. & ['Type', 'Type'] & Adding a description or explanation for an assessment. \\
\addlinespace
SP.4 Open and Type & Store process: The user accesses a resource and types information. & ['Navigation', 'Click', 'Type'] & Searching for a type of activity in the LMS. \\
\addlinespace
ShP.1 Share Resource & Share process: The user creates a resource and makes it available to others. & ['Submit'] & Creating a forum for future teacher or student use. \\
\addlinespace
ShareFlow Push with Interaction & Apply process: The user explores an automated recommendation within a task context and interacts with it. & ['push recommendation', 'click pop up window'] & Interacting with an automated recommendation to complete a task. \\
\addlinespace
ShareFlow Push without Interaction  & Apply process: The user explores an automated recommendation within a task context, but interaction is unknown. & -- & Exploring an automated recommendation, but interaction is unknown. \\
\addlinespace
Knowledge Push with User Interaction
& Apply Process: Automatic Recommendation of Organisational Knowledge (GoldMind only) & ['push recommendation', 'click pop up window'] & Teacher click and engage with the recommendation \\
\addlinespace
Knowledge Push without User Interaction & Apply Process: Automatic Recommendation of Organisational Knowledge (GoldMind only) & --- & Exploring an automated recommendation, but interaction is unknown. \\
\addlinespace
Querying & Knowledge Access: Query interaction available in GoldMind and Baseline (LLM) & Trace data when teachers interact with the query engine (e.g., Chatbot) & Teachers asking for information to support their tasks excecution \\
\bottomrule
\caption{Coding Scheme for Epistemic Network Analysis (ENA) mapping Knowledge Management sub-processes}
    \label{tab:ENAcodes}
\end{longtable}



%% file: sections/5-LessonsLearnt.tex
\section{Lessons Learnt}

This section presents the results from the three iterations of the GoldMind system development an evaluation. The results per iteration are organised around three main dimensions: \textbf{technological}, \textbf{design}, and \textbf{human factors}. At the end of each iteration, we highlight key lessons learned and their contributions to the field of Human-Computer Interaction (HCI). Knowledge management behaviours, which fall under human factors, are reported in a separate subsection.

\subsection{Iteration 1 (Month 6): Initial Evaluation and Benchmarking}

Table \ref{tab:resultsiteration1} summarises the evaluation metrics from the first iteration, comparing GoldMind with the Baseline. The results are organised and explained according to the three evaluation factors. Metric abbreviations (e.g., \circledH{m1}) are provided in the summary table for reference.


\begin{table}[ht]
\centering
\small 
\caption{Iteration 1:  Comparison of GoldMind and Baseline performance using descriptive and statistical analyses, including 95\% confidence intervals (CI) on the difference between GoldMind and Baseline and measures of improvement. For the NASA metrics, the abbreviations stand for: Mental demand (Md), Physical demand (Pd), Temporal demand (Td), Performance (P), Effort (E), and Frustration (F).}
\label{tab:resultsiteration1}
\begin{tabular}{|p{1.8cm}|c|p{2cm}|p{2cm}|p{3.8cm}|p{1.8cm}|p{1.8cm}|}
\hline
\textbf{Metric} & \textbf{Fac.} & \textbf{GoldMind} & \textbf{Baseline} & \textbf{Statistical Analysis} & \textbf{CI (difference)} & \textbf{Improvement} \\
\hline
\circledH{m1} Retrieval speed & \circledF{T} & 39 secs, SD = 108.52 & 146 secs, SD = 112.69 & $\beta =-98.99, \textbf{p = 0.000}, t =-4.11$ & -146.21 -- -51.78 & 73\% faster \\
\hline
\circledH{m2} Time to action & \circledF{H} &  32.53 secs, SD = 41.19 & 70.64 secs, SD = 61.46 & $\beta= 38.63 , \textbf{p = 0.027}, t = -2.22$ & 48.974 -- 124.96 & 35 secs faster \\
\hline
\circledH{m3} Relevant response ratio & \circledF{H} & 2.77 & 1.86 & $\beta= 0.098, p = 0.75, t = 0.51$ & -0.56 -- 0.95 & 48\% more relevant responses \\
\hline
\circledH{m4} Task duration & \circledF{H} & 717 secs, SD = 464.8 & 1172 secs, SD = 479 & $\beta = -485.76, \textbf{p = 0.000}, t = -3.80$ & -64.05 -- 285.71 & 1.68\% faster \\
\hline
\circledH{m5} Knowledge capture efficiency & \circledF{T} & 23.32 secs, SD = 39.79 & 39.89 secs, SD = 46.37 & $\beta  = -22.68, \textbf{p = 0.040}, t  = -2.05$ & -44.283 -- -1.077 & 41\% lower effort \\
\hline
\circledH{m6} Just Enough Response & \circledF{T} & 0.137, SD=0.56 & 0.20, DS=0.44 & $\beta  =-38.63, \textbf{p = 0.027}, t  = -2.21$ & -72.76 -- -4.50 &  -44\% fewer relevant responses \\
\hline
\circledH{m7} Task Completion ratio & \circledF{H} & 0.92, SD=0.89 & 0.85, SD=91 & $\beta  = 0.22, p = 1, t  = -0.36 $ & 0.50 -- 0.80 & 6\% increase \\
\hline
\circledH{m8} Task completion quality [0–1] & \circledF{H} &  0.96, SD = 0.10 & 0.97, SD = 0.10 & $\beta=0.22, p = 0.37, t = -0.09$ & -0.080 -- 0.055 & -1.9\% worse \\
\hline
\circledH{m9} NASA & \circledF{H} & Md: 2.60, SD = 0.97 \newline Pd: 2.26, SD = 1.10 \newline Td: 2.55, SD = 1.15 \newline P: 4.42, SD = 0.88 \newline E: 2.44, SD = 0.97 \newline F: 2.23, SD = 1.12 & Md: 2.63, SD = 1.01 \newline Pd: 2.25, SD = 1.13 \newline Td: 2.58, SD = 1.20 \newline P: 4.30, SD = 0.98 \newline E: 2.44, SD = 1.05 \newline F: 2.27, SD = 1.18 & Md: $\beta=-0.51, p= 0.14, t= -1.47$  \newline Pd: $\beta =-0.20, p=0.62, t= -0.48$ \newline Td: $\beta =-0.72, p=0.08, t= -1.74$  \newline P: $\beta =0.46, p=0.17, t= 1.36$ \newline E: $\beta =-0.46, p=0.23, t= -1.20$ \newline F: $\beta =-0.78, \textbf{p=0.038}, t= -2.07$ & Md: -1.19 -- 0.16 \newline Pd: -1.02 -- 0.61 \newline Td: -1.52 -- 0.08  \newline P: -0.20 -- 1.12  \newline E: -1.20 -- 0.28  \newline F: -1.52 -- -0.04 & -1.1\% less Md \newline 0.4\% more Pd \newline -1\% less Td  \newline 0.2\% more P \newline same E  \newline -1 \% less F  \\
\hline
\circledH{m10} SUS & \circledF{D} & 59.3, $SD=22.67$ & 37.8, $SD=20.30$ & $\beta = 22.84$, $\textbf{p = 0.048}$, $t = 2.20$ & 0.25 -- 45.44 & 16\% higher \\
\hline
\end{tabular}
\end{table}


\textbf{Technological Factors:} The results indicated that GoldMind outperformed the Baseline in several key aspects related to knowledge capture and dissemination. In terms of dissemination, the retrieval speed \circledH{m1}, referring to the query response time, improved significantly ($p=0.000$, $t=-4.11$) with GoldMind, which was 73\% faster. Regarding knowledge capture efficiency ($p=0.04$, $t=-2.05$) \circledH{m5}, the effort required to record knowledge was lower in GoldMind compared to Baseline. Specifically, participants using GoldMind required 41\% less effort to capture knowledge in the system (e.g., capturing and saving video tutorials). However, GoldMind underperformed in the Just Enough Response ($p=0.027$, $t=-2.21$) \circledH{m6} metric (-44\%), indicating that the responses were not always sufficiently relevant to support task completion.

\begin{figure}
    \centering   \includegraphics[width=1\linewidth]{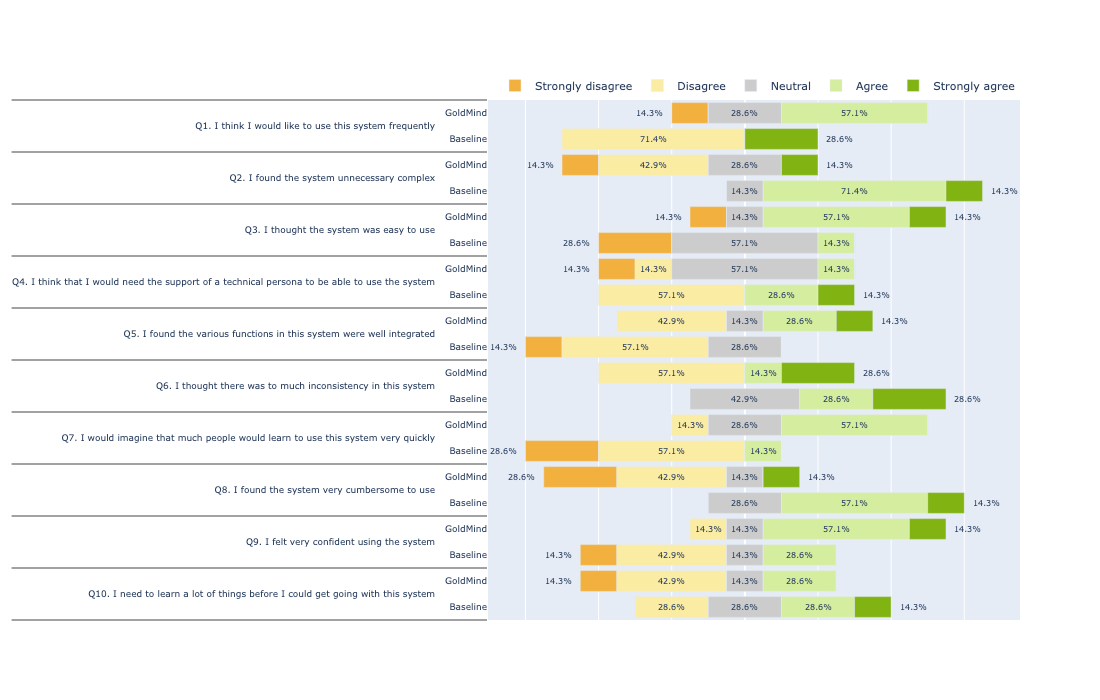}
    \caption{System Usability survey response rating by participants in Iteration 1}
    \label{fig:M6-SUS}
\end{figure}

\textbf{Design Factors} Early usability testing revealed barriers to adoption, emphasising the need for greater stakeholder involvement in system refinement. System Usability Scale (SUS) ($p = 0.048$, $t = 2.20$) \circledH{m10} scores showed a significant improvement with GoldMind (59/100) compared to the Baseline (38/100). See Figure \ref{fig:M6-SUS}.
This is in line with the task duration metric that was shorter in GoldMind than the Baseline.

\begin{figure}
    \centering
    \includegraphics[width=1\linewidth]{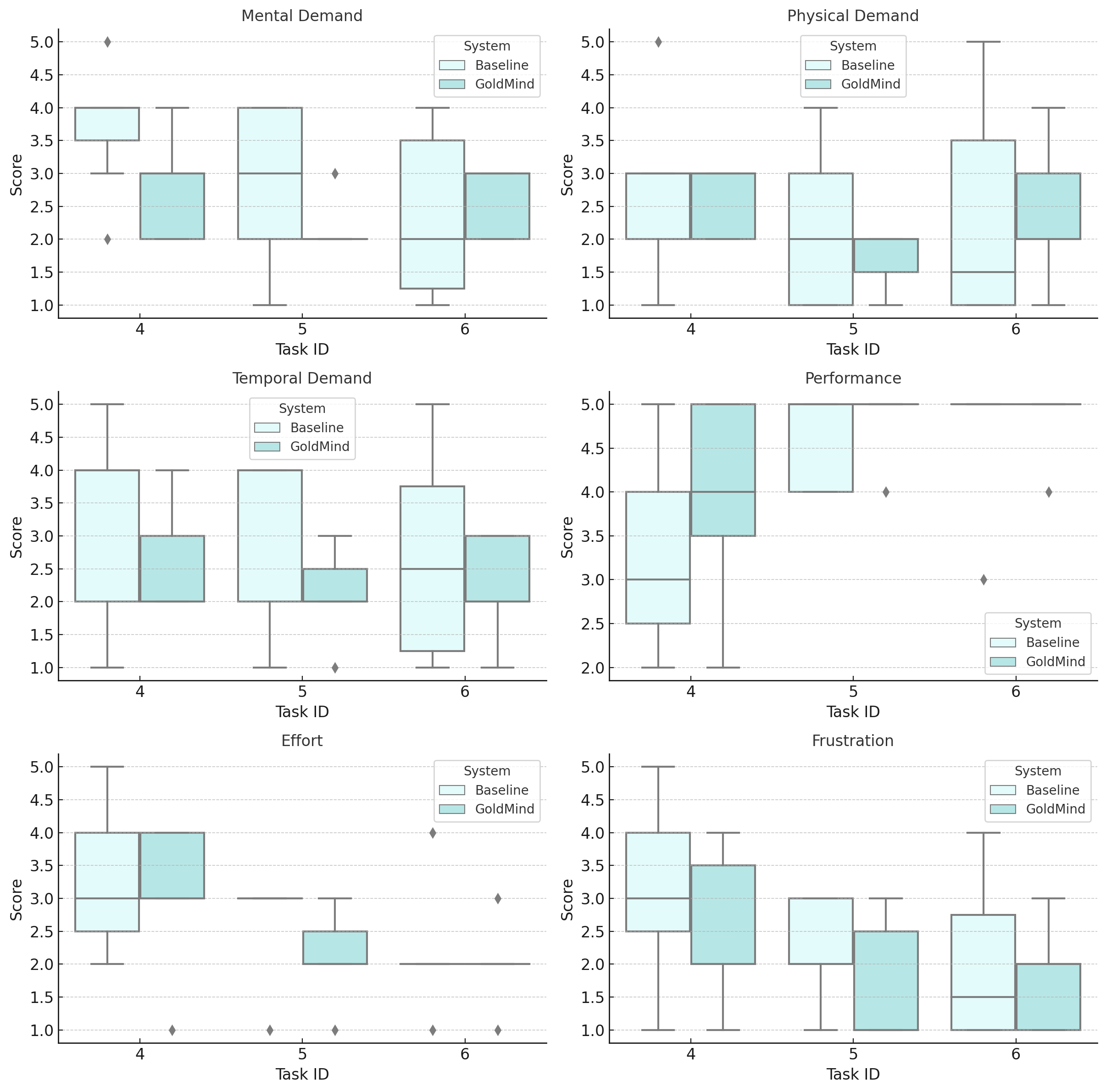}
    \caption{The results of Iteration 1 (M6) -- the descriptive statistics of the responses to the NASA task load index survey}
    \label{fig:M6-NASA}
\end{figure}

\textbf{Human Factors:} The time to action \circledH{m2} metric indicated that GoldMind users were faster (32 secs) in deciding which next action to execute ($p=0.027$, $t=-2.22$). This suggests that when responses were relevant to the task context, they also provided clear guidance on the next steps to be executed.
Task duration ($p=0.00$, $t=-3.80$) \circledH{m4} was shorter for GoldMind users compared to those using the Baseline.
The Relevant Response Ratio \circledH{m3} showed that participants found GoldMind’s responses to be more relevant (manually rated on a 0–4 scale).
The Task Completion Rate \circledH{m7} was higher for GoldMind users, showing an improvement over the Baseline.
Similarly, Task Completion Quality \circledH{m8} was also higher for GoldMind users.
Regarding the NASA Task Load Index (NASA-TLX) \circledH{m9}, the results showed that GoldMind outperformed the Baseline in five out of six usability dimensions, with the exception of Q2 – Physical Demand. Users reported feeling less rushed, experiencing lower mental demand, and feeling more confident in completing their tasks when using GoldMind. Figure \ref{fig:M6-NASA} presents the detailed NASA-TLX survey responses. Among all dimensions, only the frustration item showed a statistically significant difference ($p = 0.038$, $t = 2.20$).

\textbf{What we learnt from this iteration:} 

\textbf{Lesson 1.} \textit{In-context Knowledge Capture and Unified Access Improve Efficiency and Task Support}. 
GoldMind’s integrated features — such as the Knowledge Repository \circled{A} and Knowledge Capture in the Flow \circled{B} — effectively addressed key limitations of the baseline system, including disconnected and decontextualised knowledge. These features enabled teachers to capture and retrieve information directly within their workflow, leading to more efficient retrieval (\circledH{m1}), improved task execution (\circledH{m4}, \circledH{m7}), and higher task quality (\circledH{m8}).

\textbf{Lesson 2.} \textit{Context-aware Retrieval Requires Further Refinement}. 
Although GoldMind improved performance across several metrics, it underperformed in delivering consistently relevant responses within specific task contexts (\circledH{m6}). This suggests a need to refine the contextual retrieval mechanisms or explore alternative strategies for aligning recommendations with users' immediate task needs.

\textbf{Lesson 3.} \textit{Alignment with Teachers’ Workflows and Expectations Needs Improvement}. 
GoldMind received lower usability scores (SUS) \circledH{10} and was rated lower on human factors \circledH{m9}, particularly in terms of users’ perceived performance. These results indicate a need to enhance the system’s user experience and better align its functionality with diverse teaching workflows and expectations.

\subsection{Iteration 2: Co-Design and Iterative Prototyping}

The second iteration consisted of a co-design session and the evaluation of the systems informed by the co-design. The lessons learned on each of these stages are detailed below. 

\subsubsection{Co-design (Month 7)} \label{sec:co-design} Building on the findings from the first tool evaluation in this iteration, six co-design sessions were conducted with six teaching teams (13 teachers) to identify their needs and further refine the system's functionalities. In these sessions, the teams described their teaching workflows, evaluated the existing tools used in their practices, and, using inspirational cards, articulated the technological support they require to enhance their Knowledge Management practices. Detailed descriptions of the activities of the co-design sessions are provided in the work by Fernandez-Nieto et al. \cite{Fernandez-Nieto2024}. 

\textbf{Design Factors}: The co-design sessions led to the development of four core \textbf{design requirements} (DRs) for Knowledge Management Systems:

\begin{itemize}
    \item[\textbf{DR1}] Facilitate access to relevant knowledge. Educators need a single point to access and retrieve existing knowledge efficiently, as current repositories are scattered across multiple platforms. 
    \item[\textbf{DR2}] Granular Knowledge Storage: Educators benefit from multimodal document management that enables structured storage and retrieval of knowledge at varying levels of detail. Features like content summarisation, annotations, and metadata can enhance usability and facilitate knowledge application.
    \item[\textbf{DR3}] Proactive Recommendations and contextualised suggestions. Educators require personalised, context-aware recommendations integrated into their workflows to support knowledge application. 
    \item[\textbf{DR4}] Facilitate knowledge Sharing and Expert identification: Educators seek tools to locate and connect with community knowledge and expertise, fostering collaboration and reducing knowledge silos. Social software can enhance knowledge exchange, supporting communities of practice in higher education.
\end{itemize}

\textbf{Technological factors}:  
These design requirements (DRs) informed the design of the GoldMind system, helping identify which functionalities from Iteration 1 aligned with teachers' needs and guiding the design and implementation of new features in Iterations 2 and 3. The GoldMind system integrates four key functionalities (F1–F4).

\begin{itemize}
    \item[\textbf{F1}] Aggregating distributed knowledge: To address DR1, this functionality indexes knowledge from various repositories (e.g., university documents and videos) for easier retrieval. In Iteration 1, we introduced the Knowledge Repository (Section \ref{sec:SystemIteration1}, \circled{A}) and, based on its alignment with co-design session needs, retained it in later iterations. 

    \item[\textbf{F2}] Capturing and linking multimodal knowledge: Addressing DR2, this functionality supports knowledge capture and storage at multiple levels of granularity. In Iteration 1 (Section \ref{sec:SystemIteration1}, \circled{B}), the \textit{Knowledge Capture in the Flow} feature was used to highlight and annotate documents within their workflows. In Iteration 2 (Section \ref{sec:SystemIteration2}, \circled{C}), we extended this to video annotation and introduced \textit{ShareFlows}, a feature that captures expert interactions and translates them into reusable step-by-step workflows (Section \ref{sec:SystemIteration2}, \circled{D}). 

    \item[\textbf{F3}] Context-aware recommendations: Addressing DR3 and DR4, this functionality delivers relevant knowledge based on users’ workflows and tasks. In Iteration 2 (Section \ref{sec:SystemIteration2}, \circled{E}), we implemented a contextual recommendation system. In Iteration 3 (Section \ref{sec:SystemIteration3}, \circled{G}), we refined it by incorporating additional context, such as the timing within the semester, to improve relevance and effectiveness.  

    \item[\textbf{F4}] Knowledge sharing during workflows: Supporting DR4, this functionality enables knowledge exchange through GoldMind’s browser extension and web-clipping features, similar to \textit{Zotero} and \textit{Evernote}. Introduced in Iteration 1 (Section \ref{sec:SystemIteration1}), it was retained in later iterations based on co-design feedback. In Iteration 3, automatic recommendations were added, triggered by teachers’ workflows and task relevance (Section \ref{sec:SystemIteration3}, \circled{F}, \circled{G}). 
\end{itemize}

\textbf{Human Factors} The co-design sessions provided the guidance for the implemented functionalities. This iteration informed functionalities from the user needs derived from the DR.

\subsubsection{System Evaluation (Month 12)}

The functionalities described above (F1-4) were implemented for this subsequent evaluation in line with the DR. Table \ref{tab:resultsiteration2} summarises the evaluation metrics from the second iteration, comparing GoldMind with the Baseline. The results are organised and explained according to the three evaluation factors. Metric abbreviations for this iteration are provided in the summary table for reference.

\begin{table}[ht]
\centering
\small 
\caption{Iteration 2: Comparison of GoldMind and Baseline performance using descriptive and statistical analyses, including 95\% confidence intervals (CI) on the difference between GoldMind and Baseline and measures of improvement. For the NASA metrics, the abbreviations stand for: Mental demand (Md), Physical demand (Pd), Temporal demand (Td), Performance (P), Effort (E), and Frustration (F).}
\label{tab:resultsiteration2}
\begin{tabular}
{|p{1.8cm}|c|p{2cm}|p{2cm}|p{3.8cm}|p{1.8cm}|p{1.8cm}|}
\hline
\textbf{Metric} & \textbf{Fac.} & \textbf{GoldMind} & \textbf{Baseline} & \textbf{Statistical Analysis} & \textbf{CI (difference)} & \textbf{Improvement} \\
\hline
\circledH{m1} Retrieval speed & \circledF{T} & 1.96 secs, SD = 7.07 & 4.24 secs, SD = 7.01 & $\beta = -2.62, \textbf{p = 0.000}, t = -6.19$ & -3.45 -- -1.79 & 53.84\% faster \\
\hline
\circledH{m2} Relevant response ratio & \circledF{H} & 0.62 & 0.38 & $\beta = 3.28, \textbf{p = 0.000}, t = 6.31 $ & 0.10 --0.22 & 60.85\% more relevant \\
\hline
\circledH{m3} Task duration \newline Knowledge Capture (KC) \newline Knowledge Dissemination (KD) & \circledF{H} & KC: \newline 560.47 secs, SD = 379.33 \newline KD: \newline 597.16 secs, SD = 315.51 & KC: \newline 618.05 secs, SD = 367.65 \newline KD: \newline 759.91 secs, SD = 317.20 & KC: $\beta = -57.58, p = 0.40, t = -0.84$ \newline KD: $ \beta = -148.266, \textbf{p = 0.009}, t = - 2.62$ & KC: -191.66 -- 76.49 \newline KD: -259.13 -- -37.40 & KC: 9.3\% faster \newline KD: 21\% faster \\
\hline
\circledH{m4} Knowledge capture efficiency \newline [low - high] & \circledF{T} & 1: 49.65 secs, SD = 60.56 \newline 2: 49.39 secs, SD = 60.56 \newline 3: 43.42 secs, SD = 62.26 & 1: 80.15 secs, SD = 62.93 \newline 2: 86.16 secs, SD = 63.47 \newline 3: 52.47 secs, SD = 63.88 & $\beta = -26.74, \textbf{p = 0.000}, t = -3.94$ & -40.04 -- -13.44 & 1: 38.05\% reduction - low \newline 2: 42.67\% - mid \newline 3: 17.24\% - high \\
\hline
\circledH{m5} Just Enough Response & \circledF{T} & 0.29, $SD=0.29$ & 0.08, $SD =0.17$  & $\beta = 1.928, \textbf{p = 0.000}, t = 7.29 $ & 1.41 -- 2.44 & 273\% more useful responses \\
\hline
\circledH{m6} Task Completion ratio & \circledF{H} & 0.98, $SD=0.27$ & 0.93, $SD=0.29$ & $\beta = 0.070, p = 0.053, t = 1.93$ & -0.01 -- 0.11 & 5.35\% improvement  \\
\hline
\circledH{m7} Task completion quality [0–1] & \circledF{H} & 0.91 score, SD = 0.22 & 0.85 score, SD = 0.21 & $\beta = 0.050, p = 0.144, t = 0.52$ & -0.001 -- 0.140 & 6.85\% slightly better \\
\hline
\circledH{m8} NASA & \circledF{H}
& 
Md: 2.56, SD = 1.26 \newline 
Pd: 1.84, SD = 1.14 \newline 
Td: 2.28, SD = 1.04\newline 
P: 3.95, SD = 1.05 \newline 
E: 2.61, SD = 1.1 \newline 
F: 2.21, SD = 1.21 
& 
Md: 2.58, SD = 1.26 \newline 
Pd: 1.88, SD = 1.15 \newline 
Td: 2.30, SD = 1.05 \newline 
P: 3.92, SD = 1.05 \newline 
E: 2.62, SD = 1.1 \newline 
F: 2.23, SD = 1.22 
& 
Md: $\beta=-0.59, \textbf{p= 0.000}, t= -3.85$  \newline 
Pd: $\beta=-0.28, \textbf{p=0.009}, t= -2.59$ \newline 
Td: $\beta=-0.35, \textbf{p=0.001}, t= -2.53$  \newline 
P: $\beta = 0.54, \textbf{p=0.001}, t= 3.32$ \newline 
E: $\beta=-1.15, \textbf{p=0.000}, t= -8.20$ \newline 
F: $\beta=-1.06, \textbf{p=0.000}, t= -6.15$ 
& 
Md: -0.89 -- -0.29 \newline 
Pd: -0.49 -- -0.07 \newline 
Td: 0.22 --	0.85  \newline 
P: -0.20 -- 1.12  \newline
E: -1.42 -- -0.87  \newline 
F: -1.40 -- -0.72 
& 
-0.07\% less Md \newline 
-0.2\% less Pd \newline 
- 0.08\% less Td \newline 
0.07\% more P \newline 
- 0.03\% less E  \newline 
- 0.03\% less F \\
\hline
\circledH{m9} SUS & \circledF{D} & 85.13, $SD=24.95$ & 41.13, $SD=24.98$ & $\beta =44.00, \textbf{p= 0.000}, t=12.32$ & 37.10 -- 50.89 & 100\% more usable \\
\hline
\end{tabular}%
\end{table}


\textbf{Technological Factors:} In this second iteration, GoldMind also demonstrated significant improvements over the Baseline in knowledge retrieval efficiency ($p=0.000$, $t=-6.19$) \circledH{m1}.
For Knowledge Capture Efficiency \circledH{m4}, aligned with DR2, this iteration evaluated the capture capabilities using different levels of information complexity (low, medium, high). 
GoldMind outperformed the Baseline across all complexity levels ($p=0.00$, $t=-3.94$). Although the improvement for capturing high-complexity information was modest, the gains for low and medium complexities were substantial.

Users also received more relevant recommendations aligned with the task context, as indicated by Just Enough responses ($p=0.00$, $t=7.29$) \circledH{m5}.





\textbf{Design Factors:} System usability ($p=0.000$, $t=12.32$) \circledH{m9} improved significantly, with GoldMind scoring 85.13/100, compared to the Baseline score of 41.13/100 (a 106.99\% increase). This score also exceeded GoldMind's performance in Iteration 1 (59), indicating that the design improvements informed by the co-design sessions were effective.

\textbf{Human Factors:} The proportion of useful responses ($p=0.000$, $t=6.31$) \circledH{m2} increased from 38\% in the Baseline condition to 62\% with GoldMind. Relevant response ration motivated by DR4, were evaluated based on the teacher's level of expertise. For this analysis, participants were grouped as experts, novices, or 
journeyman. In the Baseline condition, the ratio of relevant responses was 36.63\%, 36.54\%, and 41.51\% respectively. Responses metric, which improved from 8\% in the Baseline to 29\% in GoldMind, representing a 273\% improvement. 
In contrast, the GoldMind condition showed significantly higher ratios -- 62.16\%, 62.42\%, and 61.08\% -- with lower variance. 
These results indicate that GoldMind not only generated more relevant responses but also adapted more consistently across different expertise levels compared to the Baseline. This was an improvement compared to the previous iteration.

Task duration \circledH{m3} showed improvements, with participants completing capture-focused tasks more efficiently and performing dissemination tasks significantly faster under the GoldMind condition. Tasks oriented toward evaluating dissemination (KD) were completed significantly faster with GoldMind compared to the Baseline ($p = 0.009$, $t = -2.62$). Additionally, participants completed more tasks when using GoldMind \circledH{m6}. Task quality \circledH{m7} also improved, with scores rising from 0.85 in the Baseline to 0.91 with GoldMind, indicating a 6.85\% improvement and suggesting that GoldMind facilitated slightly higher-quality task outcomes.

\begin{figure}
    \centering
    \includegraphics[width=1\linewidth]{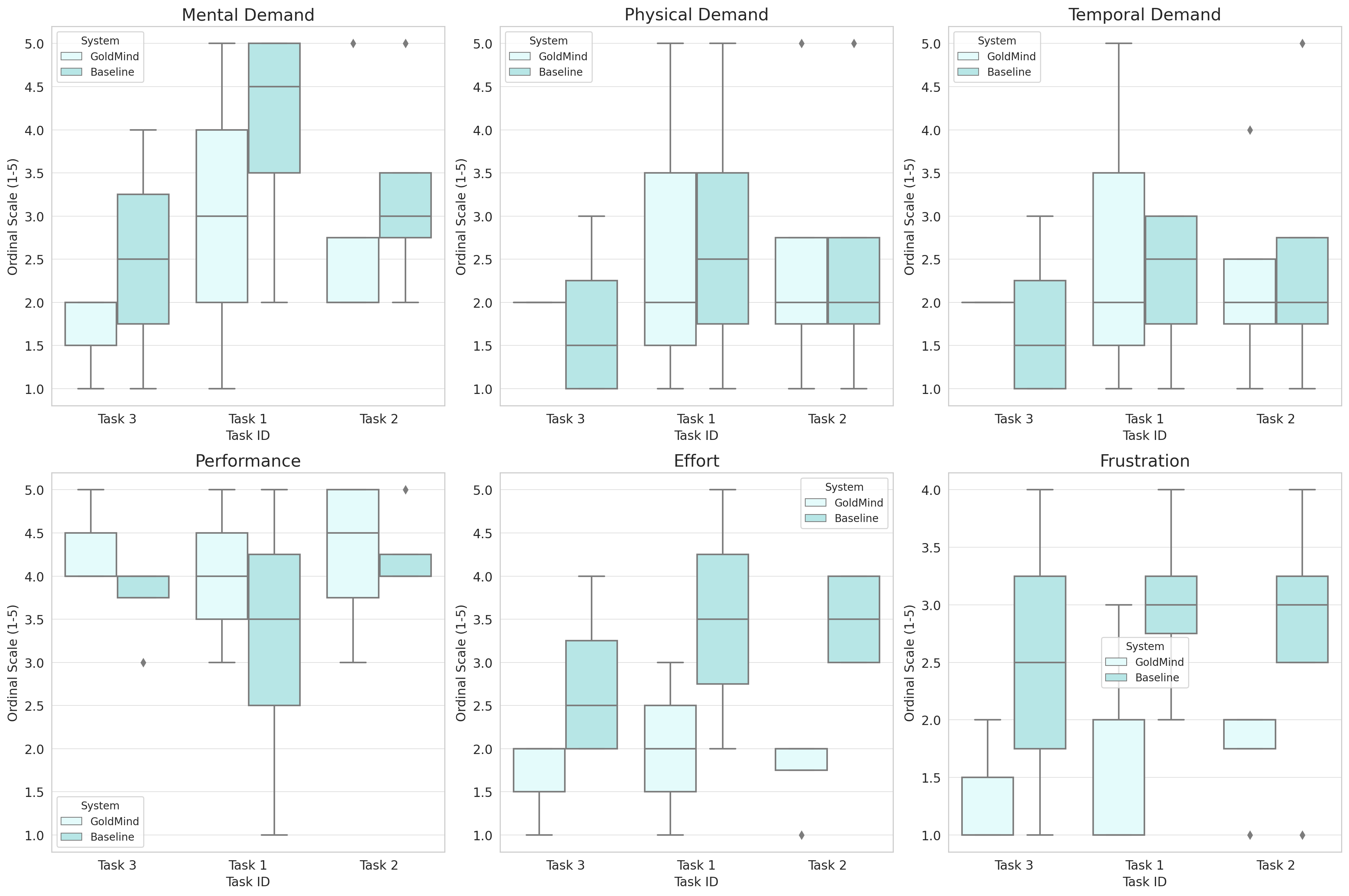}
    \caption{Results of Iteration 2 (M12) - Descriptive statistics of the NASA task load index survey responses}
    \label{fig:M12-NASA}
\end{figure}

In terms of cognitive load reduction \circledH{m8}, users reported requiring less mental effort ($p=0.000$, $t=-3.85$),  less effort to complete knowledge management tasks ($p=0.000$, $t=-8.20$), experiencing less frustration ($p=0.000$, $t=-6.15$), and perceiving improved performance when using GoldMind ($p=0.001$, $t=3.32$). Figure \ref{fig:M12-NASA} presents the responses for each question in the NASA survey. 

\textbf{What we learnt from this iteration:} 
\textbf{Lesson 4}: \textit{Co-Design Enhances Alignment with Teachers’ Needs}. The active involvement of teachers in co-design and evaluation sessions was essential for tailoring GoldMind to the real-world needs of educational teams. This collaborative approach enabled: (i) a deeper understanding of specific requirements, (ii) meaningful refinements to teachers’ interactions with the system, and (iii) the development of new features that enhanced both usability and pedagogical relevance. These improvements are reflected in the significantly higher SUS scores (\circledH{m9}), indicating better user experience and acceptance. Furthermore, a clearer understanding of teachers' workflows provided valuable insights into how passive data capture (e.g., trace data) could be used to contextualise and inform the recommendation process, demonstrated by a notable improvement in the relevance of responses for task completion (\circledH{m5}).

\textbf{Lesson 5}: \textit{Reducing Cognitive Load Enhances User Satisfaction and Task Performance}. The integration of AI-powered support features in GoldMind reduced users’ cognitive load by streamlining information retrieval (\circledH{m1}) and simplifying task execution (\circledH{m3}); see Section \ref{sec:features}, feature \circled{E}. NASA-TLX results (\circledH{m8}) indicated lower mental demand, reduced effort and frustration, and improved perceptions of task performance. These outcomes highlight GoldMind’s effectiveness in helping educators manage complex teaching tasks with less mental strain and greater confidence.

\textbf{Lesson 6}: \textit{Co-Design Informs the Need for Capturing and Sharing Expert Knowledge to Support Retention and Transfer.} The co-design sessions provided valuable insights into the direction needed to support teachers’ knowledge management practices. These collaborative discussions guided the research towards addressing these needs. Building on the existing \textit{Knowledge Capture in the Flow} capabilities (Section \ref{sec:features}, \circled{C}), the team iterated on alternative ways to meet these requirements. One key outcome was the conceptualisation of recording expert teachers’ workflows and transforming them into ShareFlows, structured, accessible representations designed to support novice users (Section \ref{sec:features}, \circled{D}). This design directly addressed the critical need for enhancing knowledge retention and transfer within educational teams, as identified in the co-design sessions. The concept of ShareFlows was further evaluated in subsequent iterations and reported in Sha et al. \cite{ShaIUI2025}. 

\begin{table}[ht]
    \centering
     \caption{Results of Iteration 2 interviews (M18). Teachers' perceptions on GoldMind and Baseline (LLM)}
    \label{tab:M18Perceptions}
    \renewcommand{\arraystretch}{1.5}
    \begin{tabular}{p{3cm} p{6cm} p{5.5cm}}
        \toprule
        \textbf{Question} & \textbf{GoldMind} & \textbf{Baseline} \\
        \midrule
        What features of the Baseline (LLM)/GoldMind tool did you find most useful? \circledF{T}
        & 
        \textit{"Additional knowledge and the use of Highlights, as it was displayed when I needed it."} (M18-P6). \newline
        \textit{"Notifications were useful, but the frequency should be revised."} (M18-P8).
        \newline
        \textit{"Recommendations help other people learn between different users."}(M18-P1).
        
        & \textit{"Q\&A, the chance to provide additional explanations to get support."} (M18-P2, P3). \newline
        \textit{"The possibility to ask questions to locate a resource."} (M18-P10, P9). \\
        \midrule
                How helpful was the information provided by the Baseline (LLM)/GoldMind for completing the tasks?\newline
        GoldMind: 4 positive, 1 negative   \newline 
        Baseline: 2 positive, 3 negative  \circledF{D}
        & \textit{"New roles (novices) would benefit from GoldMind tips"} (M18-P1). \newline
        \textit{"The push notifications are very useful, but some content seemed duplicated."} (M18-P5) \newline
        \textit{"Additional knowledge was helpful but I used it only once as I would like to complete tasks without much help."} (M18-P6, P4).
        & \textit{"At the beginning I got a few good responses but then I returned to Moodle and decided to complete the task by myself."} (M18-P3). \newline
        \textit{"For a single instruction, it works well (e.g., to find where a button is located), but for complex instructions, it is not quite informative."} (M18-P7). \newline
        \textit{"It was not helpful as it always provided wrong instructions."} (M18-P9).
        \\
        \midrule
        Did you encounter any challenges using the Baseline (LLM)/GoldMind tool? If so, please describe them. \circledF{H} 
        & 
        None (M18-P1, P4, P5). \newline
        \textit{"The recommendation push is a bit passive. Having multiple pushes can also be confusing. A way to highlight what needs attention would help."}(M18-P6). \newline
        \textit{"It may have some redundant instructions."}(M18-P8).    
        & 
        \textit{"From the responses, it seems the chat is not smart enough to understand what you mean; it was asking me for more context."} (M18-P2, P10). \newline 
        \textit{"It did not generate the whole instruction or response, or it generated late."}(M18-P3, P7, P9). \\
        \bottomrule
    \end{tabular}
\end{table}

\subsection{Iteration 3: Close to Real-World Deployment and Simplified Evaluation}


\begin{table}[ht]
\centering
\small
\caption{Iteration 3: Comparison between M18 and M24 Studies. GoldMind and Baseline performance using descriptive and statistical analyses, including 95\% confidence intervals (CI) on the difference between GoldMind and Baseline and measures of improvement.}
\label{tab:resultsiteration3}
\renewcommand{\arraystretch}{1.3}
\begin{tabular}
{|p{1.8cm}|c|p{2cm}|p{2cm}|p{3.8cm}|p{1.8cm}|p{1.8cm}|}
\hline
\textbf{Metric} & \textbf{Fac.} & \textbf{GoldMind} & \textbf{Baseline} & \textbf{Statistical Analysis} & \textbf{CI (difference)} & \textbf{Improvement} \\
\hline
\multicolumn{7}{|c|}{\textbf{M18 – New study design}} \\
\hline
\circledH{m1} Task Duration & \circledF{H} & 2037.9 secs, SD = 449.0 & 3261 secs, SD = 1080.4 & $\beta = -1223.1$, $\textbf{p = 0.006}$, $t = -2.77$ & -2088.3	-- -357.9 & 37\% faster \\
\hline
\circledH{m2} Reduction in KM time & \circledF{T} & 209.21 secs, SD = 96.15 & 469.24 secs, SD = 328.8 & $\beta = -0.117$, $\textbf{p = 0.000}$, $t = -3.54$ & -0.182	-- -0.053 & 55\% less KM time \\
\hline
\circledH{m3} Task completion quality [0--1] & \circledF{H} & 0.91 score, SD = 0.096 & 0.87 score, SD = 0.071 & $\beta = 0.032$, $p = 0.475$, $t = 0.714$ & -0.055 -- 0.119 & 26\% better \\
\hline
\circledH{m4} Failure rate reduction & \circledF{H} & No one failed & No one failed & -- & -- & -- \\
\hline
\multicolumn{7}{|c|}{\textbf{M24 – Main study}} \\
\hline
\circledD{m1} Task Duration & \circledF{H} & 708.6 secs, SD = 213.58 & 843.9 secs, SD = 213.29 & $\beta = -62.75$, $p = 0.054$, $t = -1.92$ & -126.62 -- 1.11 & 16\% faster \\
\hline
\circledD{m2} Reduction in KM time & \circledF{T} & 152.40 secs, SD = 139.48 & 274.55 secs, SD = 139.56 & $\beta =-111.735$, $p=0.324$, $t=-0.98$ & -333.58 -- 110.11 & 44\% less KM time
\\
\hline
\circledD{m3} Task completion quality [0--1] & \circledF{H} & 0.8 score, SD = 0.35 & 0.538 score, SD = 0.35 & $\beta = 0.27$, $\textbf{p = 0.000}$, $t = 3.6$ & 0.13 -- 0.43 & 47\% better 
\\
\hline
\circledD{m4} Failure rate reduction & \circledF{H} & 0.13, SD = 0.34 & 0.33, SD = 0.47 & $\beta = -2.33$, $\textbf{p = 0.001}$, $t = -3.28$ & -3.72  -- -0.94 & 60\% fewer failures \\
\hline
\end{tabular}
\end{table}

The introduction of user expertise (e.g., expert and novice) in Iteration 2 motivated the use of contextualised recommendations. The success of context-aware recommendations according to the user expertise highlighted the need to simplify evaluations to real-life task performance, leading to the redefinition of evaluation metrics in Phase 3. During the co-design sessions in Iteration 2 teachers provided details on real teaching tasks that they perform regularly. Motivated by these findings, the final phase focused on completing 
task identified during the co-design sessions to assess how well GoldMind supported knowledge management in authentic teaching scenarios.
Metrics were refined to directly measure efficiency and effectiveness in practical use cases. In addition, aiming to capture teachers' perceptions of the KMS to support their real-world tasks, we gathered information from interviews to understand their perceptions and to inform continual improvement. This section presents lessons learnt from two iterative tool evaluations.

\subsubsection{System Evaluation: new study design (Month 18)} This first evaluation was an initial test of a study with new metrics, features, and baseline. This new study design aimed to: evaluate knowledge management features; compared to a state-of-the-art Baseline tool (LLM-based Generative AI chatbot); and to simplify the evaluation process to validate effectiveness of our KMS against GenAI support. Table \ref{tab:resultsiteration3} summarises the evaluation metrics from the third iteration (Month 18), comparing GoldMind with the Baseline. The results are organised and explained according to the three evaluation factors. Metric abbreviations for this iteration are provided in the summary table for reference (gray colour).

\textbf{Technological Factors:} For the reduction in Knowledge Management time \circledH{m2}, GoldMind users retrieved relevant information 55\% faster than those using the Baseline ($p=000$, $t=-3.54$). This improvement is primarily attributed to GoldMind’s automated recommendations, which eliminated the need for teachers to manually query the system for support.

The technological factor is also informed by teachers’ perceptions, as shown in the first row of Table \ref{tab:M18Perceptions}. This row provides responses to the question: “What features of the Baseline (LLM)/GoldMind tool did you find most useful?” \circledH{m5}. In general, for GoldMind, teachers appreciated the highlights, context-aware notifications, and shared recommendations, although some suggested that the notification frequency could be improved. For the Baseline, teachers valued the ability to ask follow-up questions and locate specific resources, but they primarily relied on it for straightforward tasks (e.g., ask questions to locate a document).

\textbf{Design Factor:}
By integrating directly into teachers' workflows, GoldMind minimised disruption and enhanced adoption. In this iteration, teachers’ perceptions, shown in the second row of Table \ref{tab:M18Perceptions}, reflect their views on \textit{How helpful was the information provided by the Baseline (LLM)/GoldMind for completing the tasks?} \circledH{m7}.    
For GoldMind, most teachers found GoldMind helpful, particularly for novice users, with positive comments on timely push notifications and contextual support. A few noted content duplication and a preference for working independently, which may indicate additional opportunities to improve the content of the notification. For Baseline, feedback was mixed to negative. While some found it helpful for simple queries, many felt the chatbot failed to support more complex tasks or gave inaccurate information.

\textbf{Human Factors: }
Task duration ($p=0.006$, $d=1.47$) \circledH{m1} significantly shorter in GoldMind, likely due to the automated support provided, which eliminated the need for additional time spent querying the system. Task completion quality \circledH{m3} was also slightly higher for GoldMind. Notably, no participants failed to complete the tasks, which may be attributed to the relatively small sample size in this evaluation.

The human factor is also informed by teachers’ perceptions, as shown in the third row in Table \ref{tab:M18Perceptions}. This row provides responses to the question: \textit{Did you encounter any challenges using the Baseline (LLM)/GoldMind tool?} \circledH{m6}. In general, for GoldMind, most teachers reported no major issues, though some mentioned redundant content and the need for clearer prioritisation of notifications. For the Baseline, teachers frequently struggled with vague or incomplete responses, delayed replies, and lack of understanding from the chatbot. 


\begin{table}
\centering
\caption{Results of Iteration 3 interviews (M24). Teachers' perceptions on GoldMind and Baseline (LLM) systems}
\label{tab:M24Interviews}
\begin{adjustbox}{max width=\textwidth, max height=0.9\textheight}
\begin{tabular}{p{3cm} p{7cm} p{7cm}}
\toprule
\textbf{Question/Factor} & \textbf{GoldMind} & \textbf{Baseline} \\
\midrule
What features of the Baseline (LLM)/GoldMind tool did you find most useful? \circledF{T} & 
\textit{"ShareFlows provide examples I can follow."} (M24-P37, P35) \newline  
\textit{"Searching in a centralised repository was useful."} (P33) \newline  
\textit{"ShareFlows provided a visual representation of the task."} (P31, P26, P25, P24, P15, P12, P10, P6, P5, P2) \newline  
\textit{"ShareFlow provided additional information such as links relevant for the task."} (P31) \newline  
\textit{"It's kind of like talking with someone next to you, and then they show me."} (P25) \newline  
\textit{"ShareFlow push will help me know where to go to complete the task."} (P18) \newline  
\textit{"I think my natural inclination is to search."} (P17) 
& 
\textit{"The chat is quite simple to use."} (M24-P37, P36, P18) \newline  
\textit{"The chat has an active response, but I had to ask the same questions a couple of times."} (P35) \newline  
\textit{"Searching."} (P33) \newline  
\textit{"With ChatUI (LLM), I found it easy to find my previous queries."} (P17) \\
\midrule
How helpful was the information provided by the Baseline (LLM)/GoldMind for completing the tasks? \newline 
\textbf{GoldMind:} 26 positive, 3 negative, 1 unknown \newline 
\textbf{Baseline:} 7 positive, 17 negative, 6 unknown \circledF{D}
& \textit{"As long as the task has been done by previous experts, I can just follow their steps to complete the task."} (M24-P38) \newline  
\textit{"It is helpful, but for some tasks, it is not straightforward because other people’s experiences usually don’t completely align with mine."} (M24-P37, 36) \newline  
\textit{"It is very helpful as I can see every step I need in the flowchart (ShareFlow) for the task."} (M24-P36, P22) \newline  
\textit{"I can see the information I need."} (M24-P35) \newline  
\textit{"I was able to access the links I needed."} (P17) \newline  
\textit{"The timing of the recommendation needs to be considered, and the amount of information should be reviewed as it can be confusing."} (P34) \newline  
\textit{"The information here is generated by experts and is trustworthy."} (P33) \newline  
\textit{"The information is accurate, but it is not possible to search within ShareFlow."} (P32) \newline  
\textit{"GoldMind was better at guiding me to begin the task."} (P18) \newline  
\textit{"I know a lot of teaching staff struggle, and having a pop-up like that with instructions on how to do things would be extremely useful for them."} (P16) 
&
\textit{"As long as I can type my question, I was able to get some information to continue with the task."} (M24-P38) \newline  
\textit{"I believe it is helpful, but it fails in providing links to relevant resources for the task. Although easy to use, the information could be inaccurate."} (M24-P37, P32, P31, P24) \newline  
\textit{"Not so much, as I need to think about what question I need to ask, then the response is not quite useful for what I needed."} (M24-P36) \newline  
\textit{"It is useful, but it is just like any other chat."} (M24-P35) \newline  
\textit{"Useful and easy to use."} (P34, P28) \newline  
\textit{"The information here is AI-generated and is less trustworthy."} (P33, P5) \newline  
\textit{"The information provided by ChatUI (LLM) is more personalized."} (P30) \newline  
\textit{"If you do not know how to ask your questions, it is unlikely to get a good response."} (P4) 
\\
\midrule
Did you encounter any challenges using the Baseline (LLM)/GoldMind tool? \circledF{H} & 
\textit{"GoldMind has some delays in detecting the question I have."} (P30) \newline  
\textit{"GoldMind has so many functionalities that I feel confused with the UI."} (P28) \newline  
\textit{"I could not find the history of searches or navigation."} (P17, P8, P25) \newline  
\textit{"Pulling up notifications again was an issue."} (P16, P6, P11, P14, P16, P21, P23) \newline  
\textit{"When I tried to search for some work others have done, I felt like there is too much information."} (P4, P3, P20) \newline  
\textit{"Although it provided guidance, it is not completely tailored to my own workflow."} (P5, P13) \newline  
\textit{"The guidance was not doing what I was doing."} (P10)
& 
\textit{"Responses were only textual and the instruction was literally asking me to click the icon, but I did not know which one."} (P26, P25, P5, P10) \newline  
\textit{"The chat is not recalling the history and the context needs to be provided."} (P24, P9, P11) \newline  
\textit{"Only provided text, links were omitted."} (P22, P20, P21, P23) \newline  
\textit{"You are required to provide context in order to get a good response."} (P18, P17, P13) \newline  
\textit{"Although easy to use, the information it gave me was completely wrong."} (P16, P6, P14) \newline  
\textit{"Some responses were not correct."} (P4) \newline  
\textit{"There were hallucinations."} (P5) 
\\
\bottomrule
\end{tabular}
\end{adjustbox}
\end{table}

\subsubsection{System Evaluation: main study (M24)} For this iteration, we conducted a more extensive evaluation to validate the study design tested in the first evaluation of this iteration. We recruited 30 teachers to assess GoldMind's performance and usability metrics in the context of real teaching tasks compared to a new Baseline (GPT model). Table \ref{tab:resultsiteration3} summarises the evaluation metrics from the third iteration (M24), comparing GoldMind with the Baseline. The results are organised and explained according to the three evaluation factors. Metric abbreviations for this iteration are provided in the summary table for reference (purple colour).

\textbf{Technological Factors: }
Our tool demonstrated a reduction in knowledge management time \circledD{m2}. Under the Baseline (LLM) condition, participants spent an average of 111.735 seconds more searching for the information needed to complete the task.

The technological factor is also informed by teachers' perceptions, as shown in the first row in Table \ref{tab:M24Interviews}. This row provides provides responses to the question \textit{What features of the Baseline(LLM)/GoldMind tool did you find most useful} \circledD{m5}. In general, for GoldMind, ShareFlows were widely appreciated for providing visual, structured task guidance with examples, relevant links, and a “show-and-tell” experience. Users liked the push notifications and centralised search, although some preferred to search independently. For the Baseline, participants positively mentioned its simplicity and the ability to revisit past queries, as they were highly familiar with chatbot interfaces. However, users often had to repeat queries, and the features lacked the depth and visual clarity provided by GoldMind.

\textbf{Design Factors:}
Overall, for Knowledge Management usability \circledD{m7}, participants rated both GoldMind and the Baseline (LLM) tools highly for simplicity and ease of use. This suggests that adopting a knowledge tool — whether GoldMind or the Baseline — in the form of a web browser extension to support task workflows is positively received and can effectively deliver timely information. However, GoldMind outperformed ChatUI (LLM) in helping participants find accurate and trustworthy information. As a result, more teachers reported that the ShareFlow tool supported task completion. Participants also expressed greater satisfaction with GoldMind and a stronger interest in using it regularly compared to the Baseline, indicating its strong potential for adoption in practice (see additional details of the Knowledge Management SUS metrics in \cite{ShareFlows2025}.  

In this iteration teachers’ perceptions, shown in the second row of Table \ref{tab:M24Interviews}, reflects their views on \textit{How helpful was the information provided by the Baseline (LLM)/GoldMind for completing the task?} \circledD{m8}. In general, feedback for GoldMind was positive, consistent with the Month 18 evaluation. Teachers valued ShareFlows for their step-by-step guidance, visual clarity, and trustworthiness due to expert-generated content. While some noted that the experiences did not always align perfectly with their own tasks, highlighting the need to improve the precision of contextual recommendations that account for users’ expertise, most agreed that GoldMind supported task completion. For Baseline (LLM), Responses were mixed to negative. Although a few participants found it helpful for straightforward queries, many criticised its lack of context awareness, vague or inaccurate answers, and difficulty formulating effective prompts. AI-generated content was perceived as less trustworthy and often unhelpful for complex tasks.

\textbf{Human Factors:}
Task completion time \circledD{m1} was shorter with GoldMind. Task completion quality ($p=0.000$, $t=3.6$) \circledD{m3} (based on rubrics in appendix) was also higher when using the GoldMind tool compared to the Baseline (LLM). The results further highlight that there were fewer task failures \circledD{m4} with GoldMind ($p=0.001$, $t=-3.28$).

Teachers' perceptions also informed the human factors in this iteration. The insights presented in the third row of Table \ref{tab:M24Interviews} address the question: Did you encounter any challenges using the Baseline (LLM)/GoldMind tool? \circledD{m6}. For GoldMind, participants experienced UI complexity, difficulty finding notifications, and information overload. These perceptions could be linked to the fact that multiple features were evaluated, and the combination of learning to use a new tool while completing the task added complexity, making GoldMind feel difficult to navigate for some teachers. While the system provided guidance, teachers felt that it sometimes lacked personalisation and responsiveness to their individual workflows. However, it was also noted that GoldMind helped reduce cognitive effort. For the Baseline, key challenges included vague text-only responses, missing contextual memory, and hallucinated or incorrect information. Participants found it frustrating to repeatedly reframe their questions to receive meaningful answers. This was attributed to the tool’s inability to understand the task context on its own, requiring teachers to manually provide that context for each task in order to obtain useful responses.


To contribute to our understanding of human factors for Iteration 3, we examined teachers' knowledge management behaviours while completing tasks using both GoldMind and the Baseline, through Epistemic Network Analysis (ENA). These visual representations illustrate how teachers engaged in knowledge management processes during Iteration 3 (Month 24), with a focus on knowledge retrieval through automated push recommendations.

The results contrast the knowledge management behaviours between the Baseline LLM-based chatbot querying system and the GoldMind system (see Section \ref{sec:analysisIteration3} for a full explanation of codes used in the ENA representation).


\begin{figure}[ht]
    \centering
    \includegraphics[width=0.7\linewidth]{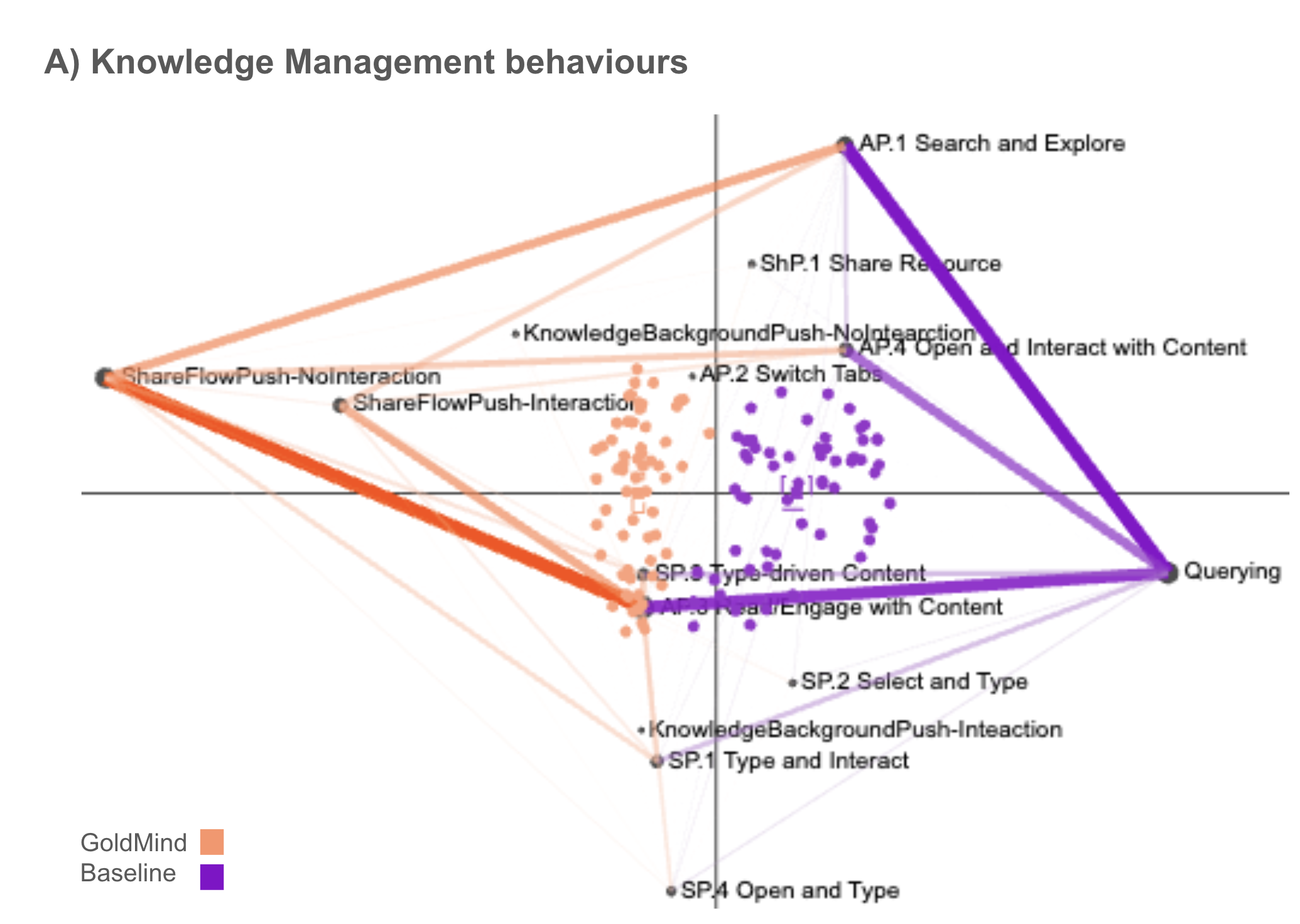}
    \caption{ENA representations of teachers' knowledge management interactions using GoldMind (light salmon), which focused on automated push recommendations, and the Baseline (dark orchid), which relied the LLM-based chatbot system for querying information. 
    }
    \label{fig:M24-Task1-2}
\end{figure}

The first dimension of the ENA space (on the right) primarily distinguished teachers who made connections among \textit{Querying} and \textit{Search and Explore} activities from those who made connections to \textit{ShareFlow Push}. This can be interpreted as a contrast between \textit{`pull'} behaviours (e.g., actively querying information) and \textit{`push'} behaviours (e.g., receiving recommendations via ShareFlow), represented on the left. This creates a clear distinction between pull (user-initiated) and push (system-initiated) knowledge management behaviours in the ENA representation.

Figure \ref{fig:M24-Task1-2} shows the subtraction between conditions across tasks. It presents teachers' knowledge management behaviours when using Baseline (LLM) and GoldMind. Thicker lines indicate more frequent connections between behaviours. For Baseline, the most common connection was from \textit{Querying} to \textit{Searching and Exploring}, suggesting teachers primarily used the Baseline chatbot to find information and then explore it. In contrast, GoldMind showed two prominent connections: from \textit{ShareFlow Push with No Interaction} to both \textit{Search and Explore} and \textit{Read/Engage with Content}.

The average ENA scores (light salmon and dark orchid), corroborate the regression analysis in Table \ref{tab:ENAregression_results}, where participants in GoldMind condition (\textit{`pull'} behaviours) are significantly farther to the left on MR1\footnote{MR refers to "means rotation", which is a dimension reduction technique that maximise the variance between conditions. In this study, we did not consider the second dimension, as the rotation method ensures the mean value of the conditions is zero in that dimension \cite{Bowman2021}} ($t= -1.255$, $\beta=-0.56$, $p= < 2e-16$, $d= 4.15$), indicating that the differences highlighted by the subtraction graph are statistically significant controlling for covariates.

\begin{table}[ht]
\centering
\begin{tabular}{|l|r|r|r|r|r|}
\hline
\textbf{Fixed Effect} & \textbf{$\beta$} & \textbf{Std. Error} & \textbf{df} & \textbf{t value} & \textbf{p-value} \\
\hline
(Intercept) & 0.328 & 0.046 & 110.75 & 7.154 & $\boldsymbol{***}$\\
SystemType (KMASS) & -0.569 & 0.025 & 89.15 & -22.769 & $\boldsymbol{***}$ \\
Task Order & -0.031 & 0.025 & 86.99 & -1.255 & 0.212 \\
Task Name (Task 2) & 0.081 & 0.025 & 86.92 & 3.282 &  $\boldsymbol{**}$ \\
Task Evaluation (Pass) & -0.044 & 0.030 & 113.42 & -1.466 & 0.145 \\
\hline
\end{tabular}
\caption{Fixed effects estimates from the linear mixed-effects model. Significance codes: *** p~<~.001, ** p~<~.01, * p~<~.05}
\label{tab:ENAregression_results}
\end{table}

\textbf{What we learnt from this iteration: } 

\textbf{Lesson 7}: \textit{Process Mining and AI-Supported Systems Improve Task Efficiency and Quality}. GoldMind significantly improved both task completion time (\circledH{m1}, \circledD{m1}) and task quality (\circledH{m3}, \circledD{m3}) compared to the baseline tools. These results suggest that integrating AI-powered knowledge management systems into teachers’ workflows can effectively support day-to-day teaching activities. In particular, by automating teacher support via LLMs (e.g., Llama) for contextual recommendations (Section \ref{sec:features}, \circled{G}) and process mining techniques to understand experts and novices workflows (Section \ref{sec:features}, \circled{F}), GoldMind enabled teachers to perform tasks more efficiently and with higher quality outcomes.

\textbf{Lesson 8}: \textit{Human-Centred Design Enables Capturing and Sharing Expert Knowledge to Support Retention and Transfer.}
The introduction of new Knowledge Capture in the Flow mechanisms (Section \ref{sec:features}, \circled{C}) in earlier iterations enabled the recording of expert teachers’ workflows and their transformation into ShareFlows (\circled{D}). By incorporating Data Storytelling principles, ShareFlows were designed to facilitate intuitive understanding and reuse of teaching practices, demonstrating the potential of representational tools for sustained knowledge sharing.

Building on these foundations, the introduction of automated ShareFlow recommendations (\circled{F}) illustrated the impact of human-centred design in providing timely and relevant support for teaching tasks (\circled{G}). This feature enabled seamless knowledge sharing within teaching teams and contributed to improved task outcomes (\circledH{m3}, \circledD{m3}) and reduced failure rates (\circledD{m4}). In addition, teachers’ perceptions of ShareFlows indicated that these representations were useful in helping them complete their tasks \circledD{m8}. These results underscore the value of designing tools that align with educators’ workflows and support collaborative knowledge retention and transfer.

\textbf{Lesson 9:} \textit{Iterative Refinement and Usability Feedback Drive Adoption}. 
The evaluation underscored the importance of iterative, user-driven co-design and usability testing in refining GoldMind. Qualitative feedback revealed the system’s strengths (e.g., usefulness and preferred features) and usability challenges, informing opportunities for further enhancement. Addressing these refinements is essential to improve user experience, increase adoption, and better support teachers in achieving their instructional goals.


\textbf{Lesson 10}: \textit{Automated KM support can encourage alternative behaviours}: The technological support provided by GoldMind prompted richer and more diverse knowledge management behaviours. This suggests that context-aware recommendations (e.g., ShareFlows push based on task context) may lead teachers to explore alternative pathways (e.g., interactions with additional resources) when completing teaching tasks.

%% file: sections/6-Discussion.tex
\section{Discussion}


In this section, we summarise the lessons learnt from our iterative study; then discuss the implications of these findings for practice, identify various limitations of our study, and suggest some potential directions for future research and development. 

\subsection{Summary of Lessons Learnt and contributions to HCI research}

Table \ref{tab:lessons_summary} summarises the 12 lessons learnt across each iteration and the ENA analysis. Based on the three main factors explored in this paper, our discussion is organised into technological, design, and human factors.

\begin{table}[ht]
\centering
\caption{Summary of Lessons Learnt and HCI Contributions Across Iterations}
\begin{tabularx}{\textwidth}{|l|X|X|}
\hline
\textbf{Iteration} & \textbf{Key Lessons Learnt} & \textbf{High-Level HCI Contributions} \\
\hline
\textbf{1} & 
\textbf{L1}. In-context knowledge capture and unified access improve efficiency and task support. (\circledH{m1}, \circledH{m4}, \circledH{m7}, \circledH{m8}) \newline
\textbf{L2}. Context-aware retrieval requires further refinement. (\circledH{m6}) \newline
\textbf{L3}. Alignment with teachers’ workflows and expectations needs Improvement (\circledH{m9}, SUS) &
- Embedded KM tools in workflows enhance efficiency, specially information retrieval for this iteration. \newline
- Early identification of user experience (UX) misfits aids design alignment \\
\hline
\textbf{2} & 
\textbf{L4}. Co-design enhances alignment with teachers’ needs. (\circledH{m5}, \circledH{m9}) \newline
\textbf{L5}. Reducing cognitive load enhances user satisfaction and task performance. (\circledH{m1}, \circledH{m3}, \circledH{m8}) \newline
\textbf{L6}. Co-design informs the need for capturing and sharing expert knowledge to support retention and transfer. (features \circled{C}, \circled{D}) &
- Co-design improves educational tool usability \newline
- Cognitive load-aware interfaces enhance satisfaction \newline
- Visual representation tools (\textit{ShareFlow}) support expert-novice knowledge transfer \\
\hline
\textbf{3} & 
\textbf{L7}. Process mining and AI-supported systems improve Task Efficiency and Quality. (\circledH{m1}, \circledH{m3}, \circledD{m4}, \circled{F}, \circled{G}) \newline
\textbf{L8}. Human-centred design enables capturing and sharing expert knowledge to support retention and transfer. (\circledH{m3}, \circledD{m4}, \circledD{m8}, \circled{D}) \newline
\textbf{L9}. Iterative refinement and usability feedback drive adoption. &
- Process mining and AI-powered, context-aware tools enhance knowledge capture and dissemination \newline
- Workflow-capture tools drive scalable knowledge reuse \newline
- Iterative feedback loops foster adoption and continuous improvement \\
\hline
\textbf{ENA analysis} 
& \textbf{L10}. Automated KM support can encourage alternative behaviours. \newline
&
- Pattern variations by condition suggest opportunities for systems to better understand knowledge management behaviours and to detect user struggles, enabling timely and tailored support.
\\
\hline
\end{tabularx}
\label{tab:lessons_summary}
\end{table}



\subsubsection{Technological factors}
Lessons one and seven (\textbf{L1}, \textbf{L7}) advance our understanding of how technology can be effectively integrated into educational contexts, particularly through in-context knowledge capture, process mining, and AI-based task support.  This aligns with the work of Di Vaio et al. \cite{DIVAIO2021220} and Asiedu et al. \cite{Asiedu31122022}, who argue that educational contexts are inherently complex due to their knowledge-intensive and multidisciplinary nature, and that further research is needed to implement effective technological support for educators. Our findings demonstrate that these technological approaches enhance task efficiency and provide meaningful support for educators in their daily workflows.

One major gap in the literature is that technological decisions in education are often made without adequately considering teachers’ actual needs \citep{corso2009CoPProcesses, Fernandez-Nieto2024}. Our study addresses this issue by employing a participatory design approach, which was instrumental in identifying context-specific requirements for KMSs (see \textbf{L4}). In particular, challenges related to task turnover and handover processes—highlighted in the literature as a pressing concern \citep{sharma2022conceptual}, were only revealed through close engagement with authentic teaching teams. This underscores the importance of co-design practices in the development of effective educational technologies.

Another persistent issue is the failure of AI systems to meaningfully incorporate human voices, resulting in a lack of mature, intelligent systems that genuinely align with educators' needs \cite{Asiedu31122022}. This was underscored in our study during Iteration 1, where \textbf{L2} and \textbf{L3} highlighted the need for ongoing refinement in system support and understanding of teachers' workflows. This lesson, along with insights from the co-design sessions, helped us recognise that KMSs are rarely embedded in teachers’ everyday practices. Our findings suggest that mechanisms such as context-aware retrieval and in-context knowledge capture can help address this gap. Importantly, the effectiveness of context-aware recommendations depends on a deep understanding of domain-specific workflows — something best achieved through participatory and iterative design processes. The co-design activities in this study revealed the importance of supporting knowledge retrieval and capture based on specific contextual needs, as emphasised in \textbf{L8}. Our findings align with previous work \cite{brynjolfsson2025generative}, which shows that such systems can enhance both the speed and quality of task completion.

Additionally, in response to concerns about using generative AI to disseminate sensitive information \cite{alavi2024knowledge}, our study suggests that when data is aligned with specific user needs and systems offer fine-grained access control, these risks can be mitigated. Future research should explore and define privacy design principles to address the potential risks of disseminating sensitive or privileged information \cite{alavi2024knowledge}. It should also investigate how knowledge capture and sharing can be embedded within KMSs to ensure secure and context-appropriate knowledge exchange among authorised users only.


\subsubsection{Design factors}

The literature highlights the need for more integrated KMSs in education, as most existing systems focus narrowly on isolated knowledge management processes like access or storage \cite{MOOCSTAM2020}. Addressing this gap, our study evaluated a KMS that supports a broader set of knowledge management processes. \textbf{L6} underscored the value of combining knowledge capture and dissemination to close \textit{the knowledge management loop}, while \textbf{L8} demonstrated the importance of moving beyond passive storage to actively capturing and sharing expert knowledge. Through co-design with educators (\textbf{L2}), we identified the need to embed such systems into real-world workflows to support timely, relevant access to knowledge. Our findings show that educators value tools that align with their daily practices and support both task execution and professional collaboration.

Our findings underscore the value of participatory design (HCD) in shaping KMs in educational settings, consistent with calls for more inclusive and context-aware systems \cite{manesh2020knowledge}. The participatory design approach (\textbf{L4} and \textbf{L6}) enabled us to involve teachers and teaching teams throughout the iterative design and evaluation phases (\textbf{L9}), ensuring the system reflected their real-world needs and practices. (\textbf{L9}) underscore the need for continual feedback collected from usability surveys (SUS) and user interviews (Tables \ref{tab:M18Perceptions} and \ref{tab:M24Interviews}) and further confirmed that HCD methods support the development of functionalities that are more closely aligned with teachers’ expectations. These insights reinforce the need to embed end-users in the design process from the outset, not only to improve usability, but to ensure adoption and long-term value of KMSs in educational settings \cite{Sarka2019KMTI}.

\subsubsection{Human factors}

As highlighted by Sarka et al. \cite{Sarka2019KMTI}, a deeper understanding of human factors is essential for the effective design and implementation of Knowledge Management Systems (KMSs). Our iterative study addressed this gap by incorporating design considerations that accounted for users’ cognitive and behavioural responses. Specifically, \textbf{L5} revealed that system complexity can increase cognitive load (Iteration 1), ultimately reducing user satisfaction. To identify such issues, we employed NASA-TLX scores and user interviews, which provided valuable insights into unforeseen complexities that may hinder usability and reduce the effectiveness of support. This evaluation process must remain ongoing, as emphasised in \textbf{L9}; usability and user perceptions are critical human factors that should be continuously monitored to guide system refinements. The final iterations of our study underscored the need to reduce complexity in ShareFlows. This finding aligns with Wang et al. \cite{WangComicInfographic2019}, who emphasised the importance of balancing contextual information, repetition, and sequencing. Future co-design of ShareFlows will aim to avoid excessive repetition and information overload, both of which can hinder usability. 

An additional human factor of effective knowledge management is the requirement for acknowledging the authentic authors of knowledge \cite{Carroll2003, alavi2024knowledge}. Our implementation of ShareFlows (Iteration 2) addressed this by allowing expert teachers to be recognised for their contributions while making their practices accessible to novices (\textbf{L6} and \textbf{L8}). This approach supports respectful and contextually relevant knowledge sharing. 

In response to calls for more qualitative and human-centred approaches in KMS design \cite{manesh2020knowledge}, and in line with broader efforts to account for the complex interplay between human actors and technological systems \cite{Sarka2019KMTI}, our study foregrounded the role of human factors through both methodological and design choices. We employed ENA as a quantitative ethnographic tool to analyse teachers' knowledge management patterns, embedding these insights throughout the co-design and evaluation phases. This approach ensured that the system aligned with real-world educational practices and user needs, contributing to the development of usable, meaningful, and adoptable technologies. Beyond usability, our ENA-based analysis allowed us to examine how specific system features, particularly automated recommendations, influenced teaching behaviours. Lesson \textbf{L10}, 
suggest that such features may positively shape task engagement and performance, demonstrating the potential of ENA to support future investigations into how KMSs mediate and transform professional practice. By revealing how teachers interact with system features over time, the ENA analysis provided a structured understanding of behavioural patterns grounded in authentic practice. This contributes to HCD by offering an empirical basis for tailoring system functionality to real user workflows and decision-making processes.

\subsection{Technological Implications: Designing for Knowledge Capture, Context, and Transfer}

The GoldMind system demonstrates how existing human-centered AI and process mining technologies can be effectively applied to support knowledge capture and dissemination in real-world educational workflows. Across iterations, we observed that features such as institutional knowledge organisation (Knowledge repository feature \circled{A}, in Section \ref{sec:SystemIteration1}), in-flow knowledge capture (Features \circled{B} and \circled{C} in Section \ref{sec:SystemIteration2}), and automated recommendations (e.g., Automated ShareFlow Recommendations \circled{F} and  Automated Contextual Recommendation \circled{G} in Section \ref{sec:SystemIteration3}) significantly enhanced knowledge management (e.g., retrieval speed, reduction in KM time) as presented in Tables \ref{tab:resultsiteration1}, \ref{tab:resultsiteration2}, and \ref{tab:resultsiteration3}. These features improved retrieval accuracy, enabled knowledge capture in usable formats (e.g., ShareFlows feature \circled{D}), reduced the time to access and record relevant information, and improved task performance, all of which may contribute to increased teacher productivity. These findings align with recent research on the use of GenAI in the workplace \cite{brynjolfsson2025generative}. We posit that the features developed across Iterations 1 to 3 hold high promise to be similarly impactful in other domains that rely on distributed digital information and operate within structured, time-sensitive workflows (e.g., health \cite{Waddell2022}, software development teams \cite{Somboonpatarakit2024}, scientific research contexts \cite{Fok2024CHI}, legal information systems \cite{JARRAHIHuman-AIPartnership}).


A key contribution of this work is the implementation of in-the-flow knowledge capture functionalities that allow educators to annotate, summarise, and store resources without disrupting their tasks, supporting knowledge retention and transfer. ShareFlows, in particular, proved effective (see Iteration 3 results in Table \ref{tab:resultsiteration3}) for capturing teaching processes and enabling knowledge transfer when automatically recommended to teachers working on similar tasks. Knowledge transfer from experts to novices remains a persistent challenge \cite{baxter2015specialized}, and has motivated research into AI-driven solutions. For example, Rodgers et al. \cite{RODGERS2023122373} proposed an AI-powered system to support expert decision-making in Smart Grid implementation, while Ottersböck et al. \cite{OTTERSBOCK2024221} explored a self-learning assistance system to mitigate knowledge loss in organisations caused by high staff turnover. Although such approaches are promising, a persistent challenge is  users' limited trust in AI-generated recommendations  \cite{alavi2024knowledge}. In our case, ShareFlows were seen as more trustworthy because teachers could identify the expert behind the recommendation (e.g., as mentioned by P33 during the interview in Iteration 3). 

Furthermore, GoldMind’s 
knowledge repositories (e.g., integrating sources using indexing algorithms), integration of context-aware, AI-powered recommendations showed a marked improvement in user usability, 
relevance of information retrieved, and alignment with teachers’ needs across levels of expertise. This demonstrates the feasibility of building adaptive knowledge systems that cater to users’ cognitive load and task complexity in real-time educational environments. 

\subsection{Design Implications: Iterative, Participatory Approaches Lead to Usable and Trusted Tools}

Our iterative HCD approach has been instrumental in teachers’ trust in GoldMind, supporting both its 
adoption and perceived effectiveness. Participatory design sessions with teachers revealed four critical design requirements: (1) unified access to distributed knowledge, (2) support for multimodal knowledge storage, (3) contextual and proactive recommendations, and (4) tools to support peer knowledge sharing. These requirements shaped GoldMind’s evolving feature set and interface design.

The design of the visual representations of teachers' workflows, or ShareFlows (feature \circled{D} in Section \ref{sec:SystemIteration2}), as storytelling artefacts enabled educators to engage with knowledge in an intuitive and structured way. This was particularly helpful for novices, who could follow task workflows visually and step by step. Teachers consistently reported that these features made complex institutional knowledge more accessible and actionable (see interviews reported in Table \ref{tab:M24Interviews}), aligning with the foundations of data storytelling, which aim to make complex information easier to understand \cite{ShaoCHI2024, Waddell2022}. Our findings also show that automated recommendations save time in task completion and improve task quality.


Our results suggest that further work is needed in system design, particularly in balancing contextual information within the visual representations of captured knowledge. This challenge is closely tied to designing a user experience (UX) that supports teachers in completing their tasks while making effective use of the system's features. 
Given the complexity of teachers’ workflows and the rich sequences captured, this study provides lessons learned for the design and implementation of a KMS features (e.g., ShareFlows), while also highlighting opportunities for future refinements — such as improved pattern highlighting, contextual scaffolding, and strategies to address usability concerns.".


\subsection{Human factors: Understanding Behaviour, Cognitive Load, and Knowledge Practices}

Not only can human-centred design improve system adoption \cite{HCDJoseph2014, Tour2014ReDesigningKM}, but it can also reshape how users interact with, capture, and apply knowledge. Although GoldMind is still in the process of being scaled across the university, the iterative design and implementation process—as well as the findings—highlight key human factors that influence these developments.

Not only can human-centred design improve system adoption \cite{HCDJoseph2014, Tour2014ReDesigningKM}, but it can also reshape how users interact with, capture, and apply knowledge. Although GoldMind is still in the process of being scaled across the university, the lessons learned from its iterative design and implementation highlight key human factors that influence this development. The iterative development of GoldMind illustrates how evaluating AI-powered knowledge management tools through successive co-design cycles can support teachers by adapting to their emerging needs. For example, the requirement to access knowledge from multiple repositories informed the integration of artificial intelligence for indexing and retrieval. Likewise, teacher feedback emphasised the need to maintain uninterrupted task flow (as identified during the co-design sessions and reflected in design requirements DR3 and DR4 in Section \ref{sec:co-design}), which led to the optimisation of contextualised AI recommendations aligned with educators’ workflows. This iterative study underscores the importance of participatory design and real-world evaluation in HCI-driven AI research to develop solutions that are attuned to authentic educational contexts.

The ENA and trace-based analyses used to understand teachers' knowledge management processes while interacting with KMSs exemplify how these analyses can be instrumental in identifying patterns in knowledge management processes. However, these approaches could be further enriched in future research through methods such as Contextual Think-Aloud \cite{Larusdottir2024}, which has proven effective in capturing nuanced user experiences in complex digital environments, such as those involving software development teams.

\subsection{Limitations}

While our study involved 108 educators across two faculties over two years, it was conducted within a single higher education institution. Broader validation across multiple institutions and disciplines is needed to test the generalisability of our findings. Additionally, although we evaluated effectiveness using realistic digital tasks, longitudinal studies are necessary to assess how GoldMind supports ongoing knowledge retention, cultural adoption, and peer collaboration over time.

Although the evaluation was grounded in realistic tasks, it was conducted in a controlled setting. Future work should explore in-the-wild deployments that integrate seamlessly into educators’ daily teaching routines and organisational systems. 

Finally, while our ENA and interaction metrics provided valuable insights into patterns of knowledge management processes, future research should investigate the social and collaborative dimensions of knowledge management, such as peer knowledge co-construction and mentoring, to deepen our understanding of the human ecosystem surrounding KMS use. Additionally, a limitation of our ENA approach is that the coding process (outlined in Section \ref{sec:analysis}, Table \ref{tab:ENAcodes}) was not formally validated. Typically, this involves comparing human-coded data with automated coding to assess reliability. Future studies should incorporate this validation step to uncover further insights into teachers’ behaviours when using KMSs to complete their knowledge management tasks.

\subsection{Conclusion}

This paper presented a two-year human-centred study on the iterative design, implementation, and evaluation of GoldMind, a knowledge management system developed to support educators in higher education. Grounded in knowledge management processes and informed by real teaching tasks, our approach underscore technological, design, and human challenges that provide evidence for successful adoption and impact of knowledge management system  that can be extended to not only educational settings.

Through three iterations involving 108 teachers in higher education, we identified key lessons learnt -- including challenges and opportunities -- in capturing, representing, and sharing expert knowledge to support teaching workflows. Our findings demonstrate the value of integrating AI technologies such as process mining and generative AI-based summarisation within a co-designed system, enabling teachers to access timely, relevant, and context-aware information. The use of visual process representations (e.g., ShareFlow), proactive knowledge recommendations, and a focus on usability contributed to more efficient task completion and improved user satisfaction.

Critically, this work illustrates how iterative, participatory, and interactive design processes not only improve the usability and effectiveness of intelligent systems but also deepen our understanding of human knowledge management processes . The analysis of teachers' knowledge management processes revealed the importance of timing, context, and trust in shaping engagement with system recommendations and support mechanisms.

By combining human-interaction and knowledge management perspectives, our study contributes practical insights into designing knowledge management systems that are adaptable, usable, and sensitive to educators' evolving needs. Future work should expand these findings through longitudinal, multi-institutional deployments and explore collaborative and cultural aspects of knowledge sharing in greater depth.
